\documentstyle[aps,eqsecnum,preprint,floats]{revtex}
\newbox\rotbox
\def\be{\begin{eqnarray}}
\def\ee{\end{eqnarray}}

\def\GeV{\nobreak\,\mbox{GeV}}
\def\fm{\nobreak\,\mbox{fm}}

\def\qbar{\overline{q}}
\def\ubar{\overline{u}}
\def\dbar{\overline{d}}
\def\sbar{\overline{s}}

\def\bra#1{\langle #1|}
\def\ket#1{| #1\rangle}

\def\nucbrap{\bra{ p'}}   
\def\nucket{\ket{ p}}

\begin{document}
\draft

\preprint{\vbox{\null \hfill ECT*/Sept/95-04}}


\title{Dispersion Analysis of the Strange Vector Form Factors \\
of the Nucleon}

\vskip 4.5cm

\author{Hilmar Forkel}
%


\address{European Centre for Theoretical Studies in Nuclear Physics 
and Related Areas, \\
Villa Tambosi, Strada delle Tabarelle 286, I-38050 Villazzano, Italy}
%
%
\date{September 1995}

\maketitle

\vskip -0.5cm

\begin{abstract}

\vskip -0.5cm

We analyze the nucleon matrix element of the strange quark vector 
current in a nucleon--model independent dispersive approach with 
input from the current world data set for the isoscalar electromagnetic 
form factors. The update of Jaffe's minimal 3-pole ansatz for the 
spectral functions yields a 40\% larger (Sachs) strangeness radius, 
$(r_s^2)_{Sachs} = 0.20 \, {\rm fm}^2$, and a by 20\% reduced 
magnitude of the strangeness magnetic moment, $\mu_s = -0.26$. In the 
pole approximation these values are shown to be upper bounds. After
extending the ansatz in order to implement the asymptotic QCD momentum 
dependence (which the 3-pole form factors cannot reproduce), we find 
the magnitude of the 3-pole results reduced by up to a factor of 2.5. 
The signs of the leading moments originate primarily from the large 
$\phi$-meson couplings and are generic in the pole approximation.

\end{abstract}

\section{Introduction}
\label{intro}

Nonvalence quark distributions in hadrons arise from subtle and little 
understood quantum effects in the hadronic wave functions, which provide 
a unique key to hadron structure beyond the naive quark model 
\cite{der75}. The sea quark distributions of the nucleon in the 
strangeness and charm sectors are particularly interesting and much 
studied examples. While charm admixtures are mainly probed in hard 
scattering processes \cite{bro81}, strangeness fluctuations in the 
nucleon can produce (due to the smaller strange quark mass) much 
larger effects which are in some Lorentz channels directly measurable 
at low energies. Mounting experimental evidence indeed indicates 
significant values for the nucleon matrix elements of the strange scalar 
\cite{che71,don86,gas91} and axial vector \cite{ahr87,ash88,deut} 
currents.    

In order to further advance the understanding of the nucleon's 
strangeness content, both experimental and theoretical studies beyond 
these two channels are crucial. For once, the channel dependence provides 
insight into the dynamical origin of the strange-quark distribution. 
Flavor mixing instanton-induced interactions, for example, reveal 
themselves in a pronounced and characteristic channel dependence pattern 
\cite{iof91}. Furthermore, some controversial assumptions in the analysis 
of the existing data (for example in the extrapolation and small-$Q^2$ 
evolution and in the treatment of $SU(3)$ violations) can be avoided in 
other channels. 

The present paper deals with the still unmeasured vector channel, i.e. 
with the nucleon matrix element of the strange vector current. This 
matrix element is experimentally accessible at low energies and has some 
useful theoretical properties. In close analogy with and as a part of 
the electromagnetic distributions it describes the nucleon's strangeness 
charge and current distributions by Dirac and Pauli form factors. 
Furthermore, strangeness conservation renders these form factors 
scale independent (up to weak corrections), which avoids 
complications due to nonperturbative evolution from the outset. 

As already mentioned, essentially no experimental information on the 
vector form factors exists up to now (apart from a reanalysis of 
older neutrino scattering data \cite{gar93} with too poor statistics 
to be conclusive). With the present experimental techniques, however, 
they can be directly measured by parity-violating lepton scattering off 
different hadronic targets \cite{mck89,bec89,mus94}. Four experiments 
of this type are in preparation at CEBAF \cite{mus94,ce17,ce4,ce10} and 
MAMI \cite{mus94,hei93}, while SAMPLE \cite{mit} at Bates already started 
to take data. These experiments will measure for the first time sea quark 
effects in hadrons at low energies.  

In anticipation of the forthcoming data several theoretical estimates of 
the strange form factors, primarily on the basis of nucleon models, have 
appeared in the literature. Since sea quark effects arise from a delicate 
interplay of quantum effects in QCD, their reproduction in hadron models 
is much more challenging than the calculation of the standard static 
observables. Reflecting these difficulties, present nucleon model 
estimates \cite{par91,wei92,mus93,nos,hon93,for94,koe94,kim95} contain 
large and often uncontrolled theoretical uncertainties. For the Dirac form 
factor, in particular, the predictions vary by over an order of magnitude 
and in their sign. A comparison of some of these estimates can be found 
in Ref. \cite{for94}. Lattice simulations of the strange form factors 
have not yet been carried out since the computational demands increase 
substantially when quark-line disconnected contributions have to be taken 
into account (see however \cite{keh94}).

In the present paper we bypass the problems associated with these 
dynamical calculations by persuing a dispersive, nucleon--model 
independent approach. It was initiated by Jaffe \cite{jaf89} and becomes 
practicable since the isoscalar electromagnetic current carries the same 
quantum numbers as the strange current and thus couples to the nucleon 
through the same intermediate states. The available experimental data on 
the electromagnetic nucleon form factors can therefore be used as input 
for the strange form factor analysis.

After updating the minimal dispersive analysis of Jaffe with input from 
the current world data set for the electromagnetic form factors, we will 
focus on extensions of the spectral functions which implement information 
from QCD at high momentum transfers. We will be particularly interested 
in the implications of the QCD asymptotics for the low--momentum behavior 
of the form factors and their first non-vanishing moments, i.e. the 
strangeness radius and magnetic moment. A better understanding of the 
low--momentum regime is also needed for the experimental determination of 
the moments: since one has to extrapolate the data to zero momentum 
transfer, the resulting values will be sensitive to the assumed 
low--momentum dependence of the form factors \cite{for94}. 

The paper is organized as follows. In the next section we outline the 
general ideas behind the dispersive treatment of the strange form factors 
and describe their implementation in some detail. We then update Jaffe's 
minimal 3-pole estimate in Section \ref{3polup} with input from new fits 
to the electromagnetic form factors. In Section \ref{b3pol} we discuss 
several extensions of the approach in pole approximation which are 
designed to adapt the large (spacelike) momentum behavior to 
predictions from QCD counting rules. Some generic features of the 
dispersive analysis and their impact on the sign of the moments are 
pointed out in section \ref{gen}. Section \ref{sum}, finally, summarizes 
the main results and contains our conclusions. A partial summary of 
these results was reported previously in \cite{for95}.

\section{Strange vector form factors and 3-pole estimate}
\label{3pol}

In the absence of time reversal violations the nucleon matrix element 
of the strange quark vector current\footnote{Note the nonstandard sign 
convention for the strangeness charge \cite{jaf89}, which carries over 
to the sign of its hypercharge contribution.} 
\be
J_\mu^s = \bar{q} \gamma_\mu (B - Y) q = \bar{s} \gamma_\mu s, \quad 
{\rm with} \quad Y = B - S = \frac{\lambda^8}{\sqrt{3}}
\ee
is completely determined by two invariant amplitudes, the strange Dirac 
and Pauli form factors $F_{1}^{(s)}$ and $F_{2}^{(s)}$:
\be
\nucbrap  \sbar \gamma_\mu s \nucket = \overline{N}(p^\prime) 
\left(\gamma_\mu F_{1}^{(s)}(q^2) +
i\frac{\sigma_{\mu\nu}q^\nu} {2M_N} F_{2}^{(s)}(q^2)
\right) N(p). \label{sff}
\ee
(Here $q = p^\prime - p$ is the momentum transfer of the current and 
$N(p)$ denotes the nucleon spinor.) The above decomposition is analogous 
to that of the electromagnetic current matrix element. Strangeness 
conservation and the nucleon's zero overall strangeness charge imply, 
however, a different normalization, $F_{1}^{(s)}(0) = 0$, of the Dirac 
form factor. 

Alternatively, the matrix element (\ref{sff}) can be described in terms 
of the electric and magnetic Sachs form factors:  
\be
G_E^{(s)}(q^2) &=& F_1^{(s)}(q^2) + {q^2\over4M_N^2}F_2^{(s)}(q^2), 
\nonumber\\*[7.2pt] G_M^{(s)} (q^2) &=& F_1^{(s)}(q^2) + F_2^{(s)}(q^2) .
\label{sachs}
\ee
Due to their association with the strangeness charge and current 
distributions in the Breit frame the Sachs form factors have a 
somewhat more direct physical interpretation. 

The strangeness radius $r^2_s$ and the strangeness magnetic moment 
$\mu_s$ are defined as the first nonvanishing moments of either the 
Dirac/Pauli or the Sachs form factors, 
\be
r^2_{s} =  6 \frac{d }{d q^2} F_1^{(s)} (q^2) |_{q^2=0},  \qquad 
(r^2_{s})_{Sachs} =  6 \frac{d }{d q^2} G_E^{(s)}(q^2) |_{q^2=0}, \qquad
\mu_s = F_2^{(s)} (0) = G_M^{(s)}(0). \label{mur}
\ee 
Both definitions are currently in use. 

Our analysis of the strange form factors starts from the dispersion 
relations  
\be
F_i^{(s)} (q^2) = \frac{1}{\pi} \int_{s_0}^\infty d s \,
\frac{Im \{F_i^{(s)}(s)\}}{s-q^2}.
\ee
Subtraction terms are omitted since they play no role in the following 
discussion. The singularities of the form factors are located above the 
three-pion threshold (in the limit of good G-parity), i.e. at real, 
time-like $q^2 \ge s_0 = (3 m_{\pi})^2$. The spectral functions 
$\pi^{-1} Im \{F_i^{(s)} \}$ receive contributions from all on-shell 
intermediate states with $ I^G \, J^{PC} = 0^- 1^{- -}$ through which 
the strangeness current couples to the nucleon. 

Our aim in the remainder of this paper will be to construct N-pole 
approximations
\be
\frac{1}{\pi} Im \{F_i^{(s)}(s) \} = \sum_{v=1}^N B_i^{(v)} m_v^2 \,
\delta (s - m_v^2),  \qquad i \in \{1,2\},
\label{npsd}
\ee
for the spectral densities, i.e. to represent the intermediate states by 
sharp vector meson poles. This ansatz provides an excellent description 
of the two lowest--lying resonances in the isoscalar channel, the $\omega$ 
and $\phi$ mesons, with their respective widths of 8 and 4 MeV. The 
additional poles summarize strength from higher-lying resonances and from 
the multi-particle continuum.  

The $N$--pole ansatz (\ref{npsd}) contains $3 N$ a priori undetermined 
mass and coupling parameters. It is crucial for the reliability of the 
dispersion analysis that these parameters, and in particular those of 
the low-lying poles, are determined as accurately as possible. 
Fortunately, the couplings and masses of the $\omega$ and $\phi$ as well 
as the mass of the third pole can be estimated model--independently from 
experimental input, as noted by Jaffe \cite{jaf89}. This estimate  
relies on the fact that the isoscalar electromagnetic current 
$J^{(I=0)}_\mu = J^{Y}_\mu / 2$ shares the quantum numbers of the strange 
current and thus couples through the same intermediate states to the 
nucleon. 

Since the isoscalar form factors $F^{(I=0)}_i$ have been measured in a 
large momentum range ($0 \le Q^2 = - q^2 \le 30 \, {\rm GeV}$) and are 
well fitted \cite{hoe76,mer95,hoe93} by a dispersive 3-pole ansatz 
analogous to (\ref{npsd}), the masses and couplings of these three pole 
states in the strange form factors can be estimated from the fit
parameters: The three masses $m_1 - m_3$ in (\ref{npsd}) are identified 
with the pole positions found in the fits (which implies in particular 
$m_1 = m_\omega, m_2 = m_\phi$) and the strange $\omega$ and $\phi$ 
couplings\footnote{Depending on the context we will sometimes substitute 
the symbols $\omega$ and $\phi$ for the values $v=1,2$ of the pole 
index.} $B_{1,2}^{(\omega,\phi)}$ are related to the fitted values of 
the corresponding isoscalar couplings $A_{1,2}^{(\omega,\phi)}$ by 
exploiting the known flavor structure 
\be
\ket{\omega} &=& \cos\epsilon\,\frac{1}{\sqrt{2}} \left( \ket{ 
\ubar\gamma_\mu u} + \ket{\dbar\gamma_\mu d} \right) - \sin\epsilon\,
\ket{\sbar \gamma_\mu s }, \nonumber \\*[7.2pt]
\ket{\phi} &=& \sin\epsilon\,\frac{1}{\sqrt{2}} \left( 
\ket{\ubar\gamma_\mu u} + \ket{\dbar\gamma_\mu d} \right) + 
\cos\epsilon\, \ket{\sbar \gamma_\mu s}  \ , \label{states}
\ee
of the $\omega$ and $\phi$ poles. (The physical flavor wave functions 
deviate only slightly from the ideally mixed states, i.e. the mixing 
angle $\epsilon=0.053$ \cite{jain} is small.) The couplings $g(V,J)$ 
of the vector mesons $V= \omega, \phi$ to the neutral currents $J = 
J^{(I=0)}, \, J^{(s)}$ (defined via $\bra{0} \, J_\mu \, \ket{V} \, = 
g(V,J) \,\, m^2_V  \, \varepsilon_\mu$, where $\varepsilon_\mu$ is the 
polarization vector of the $V$) are determined under the assumption 
that the quark current of flavor $i$ couples with universal strength 
$\kappa$ exclusively to the component of the vector-meson wave function 
$(\qbar_j q_j)_V$ with the same flavor, i.e.
\be
\bra{0} \, \qbar_i \gamma_\mu  q_i \,\ket{\, (\qbar_j q_j)_V} \, = 
\kappa \, m^2_V \, \delta_{i j} \, \varepsilon_\mu. \label{me}
\ee
This formula reproduces the empirical values of the electromagnetic 
coupling ratios to within a few percent. For the couplings to the 
neutral currents it leads to the expressions 
\be
g(\omega,J^{I=0}) =& \frac{\kappa}{\sqrt{6}} \sin (\theta_0 + 
\epsilon), \qquad \qquad \,\, 
g(\omega,J^{s}) &= - \kappa \sin \epsilon, \nonumber \\
g(\phi,J^{I=0}) =& - \frac{\kappa}{\sqrt{6}} \cos (\theta_0 + 
\epsilon), \qquad \qquad 
g(\phi,J^{s}) \,\,\mbox{} &= \kappa \cos \epsilon,
\ee
where $\theta_0$ is the ``magic angle'' with $\sin^2 \theta_0 = 1/3$.

After parametrizing the vector--meson nucleon couplings as 
$g_i (\omega_0,N) = g_i \cos \eta_i$, $ g_i(\phi_0,N) =  g_i \sin 
\eta_i$ (the index $i = 1,2$ refers to the $\gamma_\mu$ and $\sigma_{\mu 
\nu} q^\nu$ couplings and $\omega_0, \phi_0$ denote the ideally mixed 
states), the four couplings $B_{1,2}^{(\omega, \phi)}$ can be related to 
the corresponding (fitted) isoscalar couplings $A_{1,2}^{(\omega, \phi)}$ 
(which determine phenomenological values for $\eta_i$ and $\kappa g_i$)
through the relations 
\be
A_i^{(\omega)} =& \frac{\kappa g_i}{\sqrt{6}} \sin (\theta_0 + 
\epsilon) \cos (\eta_i + \epsilon), \qquad
B_i^{(\omega)} &= - \kappa g_i \sin \epsilon \cos (\eta_i + 
\epsilon)  \label{omegcoupl}, \\
A_i^{(\phi)} =& - \frac{\kappa g_i}{\sqrt{6}} \cos (\theta_0 + 
\epsilon) \sin (\eta_i+ \epsilon), \qquad
B_i^{(\phi)} \,\, &= \kappa g_i \cos \epsilon \sin (\eta_i+ \epsilon).
\label{phicoupl}
\ee
The numerical values of the couplings $B_{1,2}^{(\omega, \phi)}$ for the 
different fits to the electromagnetic form factors (cf. Table I) are 
listed in Table II. 

The normalization of the Dirac form factor requires (in the given 
framework) at least one more pole besides the $\omega$ and the $\phi$. 
In order to complete the construction of the spectral densities 
(\ref{npsd}) for the minimal 3-pole ansatz we therefore have to 
determine two more couplings, $B_{1}^{(3)}$ and $B_{2}^{(3)}$. For 
this purpose flavor symmetry arguments offer no help since the flavor 
content of the strength associated with the third pole is unknown. 
Instead, we fix these couplings by imposing moderate constraints 
(i.e. superconvergence relations) on the asymptotic behavior of the 
form factors, 
\be
\lim_{q^2 \rightarrow \infty}  \,\, F_1^{(s)} (q^2) \rightarrow 0 
\qquad &\Rightarrow& \quad \sum_v B_1^{(v)} = 0, \label{as1} \\ 
\lim_{q^2 \rightarrow \infty} \,\, q^2 \, F_2^{(s)} (q^2) \rightarrow 0 
\qquad &\Rightarrow& \quad \sum_v B_2^{(v)} m_v^2 = 0. \label{as2}
\ee
The first of these relations is needed in any case since it also 
normalizes the Dirac form factor. Table II contains the numerical values 
of the third--pole couplings which follow from solving these constraints. 
Since the normalization condition can be used to write $F_1^{(s)}$ with 
one subtraction, the final expressions for the 3-pole form factors 
become 
\be
F_1^{(s)} (q^2) &=& q^2 \sum_{v=1}^3 \, B_1^{(v)} \, \frac{1}{m^2_v 
- q^2} ,  \label{f1s} \\
F_2^{(s)} (q^2) &=& \sum_{v=1}^3 \, B_2^{(v)} \, \frac{m^2_v}{m^2_v 
- q^2}. \label{f2s}
\ee\
In this form they were used by Jaffe \cite{jaf89}, with parameters 
extracted from three of H{\"o}hler {\it et al.}'s fits \cite{hoe76} 
(cf. Tables I and II). The resulting form factors are plotted in Figs. 
1a and 1b. Note that the overall scale of the form factors depends 
rather sensitively on which parameter set is used. For an estimate of 
the strangeness radius and magnetic moment Jaffe averaged over all 
three fits and obtained $r_s^2 = (0.16 \pm 0.06) \, {\rm fm}^2$, 
$(r_s^2)_{Sachs} = (0.14 \pm 0.07) \, {\rm fm}^2$ and $\mu_s = - (0.31 
\pm 0.09)$ (in units of the nuclear magneton). The indicated errors 
originate solely from differences between the fits and provide therefore 
at best a rough lower bound on the complete error.

The 3--pole results for the leading moments are surprisingly large, of 
the order of the neutron charge radius $r^2_n = -0.11 \, {\rm fm}^2$ 
and the nucleon's isoscalar magnetic moment $\mu^{(I=0)} = 0.44$, 
respectively. The main origin of these values can be traced to the large 
couplings of the $\phi$. As will be shown below, the dominant role of 
the $\phi$ pole in the low--$s$ region of the spectral functions 
determines together with the asymptotic constraints most of 
the momentum dependence of the form factors\footnote{The results of the 
dispersive analysis rely therefore strongly on the identification of 
the second pole in the isoscalar form factor fits with the physical 
$\phi$ meson.}. 

We conclude this section by noting that the simplicity of the 3-pole 
ansatz implies both advantages and limitations. On the one hand it 
requires a minimal number of parameters to be fixed and avoids the 
increasingly less reliable description of higher-lying strength in 
terms of additional poles. On the other hand it leads to a dipole 
behavior of the form factors (see the following section) and cannot 
accomodate the faster decays which QCD counting rules predict at large 
space-like $q^2$. This issue will be addressed in Section \ref{b3pol}.

\section{Update of the 3-pole estimate}
\label{3polup}

The 3-pole estimate of the last section was based on mass and coupling 
parameters derived from the twenty year old H{\"o}hler fits. A recently 
performed new fit to the current world data set for the electromagnetic 
form factors by Mergell, Meissner and Drechsel (MMD) \cite{mer95} 
permits an update of this analysis, which will be the main subject of 
the present section. We will also discuss some characteristic features 
of the resulting spectral densities and their impact on the momentum 
dependence of the form factors. 

The MMD fits were designed to reproduce both the asymptotic power 
behavior\footnote{One of the necessary superconvergence relations for 
this behavior was not imposed by H{\"o}hler {\it et al}.} 
\cite{bro75} of the electromagnetic form factors, $F_i \rightarrow 
q^{-2(i+1)}$, and the logarithmic QCD corrections \cite{lep80}. Since 
the asymptotic behavior arises at least partially from continuum 
contributions, a more complete description should reduce the 
continuum contamination of the extracted pole couplings. One would 
expect this effect to be strongest for the third, effective pole, and 
indeed the MMD couplings $A_i^{(3)}$ are substantially smaller than 
the average couplings $(\bar{A}_1^{(3)})_H$ of H{\"o}hler et al.:
\be
\frac{ (A_1^{(3)})_{MMD} }{ (\bar{A}_1^{(3)})_H } = 0.19, \qquad 
\frac{ (A_2^{(3)})_{MMD} }{ (\bar{A}_2^{(3)})_H } = 0.67.
\ee
The accuracy of the fitted $\omega$ and $\phi$ couplings (which 
change considerably less, cf. Table I, but are of pivotal importance 
in the determination of the corresponding strange current couplings) 
should also benefit from the improved continuum description and enhance 
the reliability of the strange form factor analysis. 

As perhaps another consequence of the improved continuum description, 
the MMD fit finds the third pole at the mass of a well--established 
resonance, the $\omega(1600)$. This strengthens the rationale for 
adopting the same $m_3$ in the strange form factors: if the third pole 
mostly summarized continuum contributions (as assumed in \cite{hoe76}), 
the response to the strange and hypercharge currents could in principle 
be centered at different invariant masses in the spectral functions. 
Still, the association with the $\omega(1600)$ should not be taken too 
seriously since the quality of the data, the ill-posed fitting problem 
\cite{sab75} and limitations of the pole ansatz do not allow a very 
accurate determination of the pole positions\footnote{H{\"o}hler {\it et 
al}. \cite{hoe76}, for example, could fit the older data with comparable 
accuracy for values of $m_3$ between 1400 and 1800 MeV.}. In the spectral 
functions of the strange form factors, furthermore, strength around $1600 
\, {\rm MeV}$ may originate from the nearby $\phi(1680)$ resonance with 
its larger strangeness content. Some support for this possibility will 
emerge in section \ref{b3pol}.

Concluding this brief disucussion of improvements in the MMD fit which
are beneficial for the strange form factor analysis, we stress that its 
probably most important new feature is the considerably expanded 
experimental data base. It consists of the current world data set for 
the electromagnetic form factors, which has grown, in particular for the 
neutron electric form factor, substantially over the last two decades 
since H{\"o}hler {\it et al}.s fits appeared.  

The parameter update for the strange 3-pole form factors follows  
essentially the procedure of the preceding section. The three pole 
masses are identified with those of the MMD fits, i.e. $m_\omega = 781 
\,{\rm MeV}, m_\phi = 1019 \,{\rm MeV}, \mbox{ and } m_3 = 1600 \, {\rm 
MeV}$. An additional step is required to extract the pole couplings,  
since the MMD couplings  
\be
A_i^{(v)} (q^2) = A_i^{(v)}  \, \frac{L(q^2)}{L(m_v^2)},\quad \qquad 
L(q^2) = \left[\ln \left( \frac{9.733-q^2}{0.350} \right) 
\right]^{-2.148} \label{momcoup}
\ee
(in the ``multiplicative'' parametrization of Ref. \cite{mer95}) have an 
effective momentum dependence devised to reproduce the logarithmic QCD 
corrections in the asymptotic region. However, since the $A_i^{(v)}(q^2)$ 
vary by less than $ 10 \%$ in the low-momentum region $0 \leq Q^2=-q^2 
\leq 2 \, {\rm GeV}^2$ and hardly affect the momentum dependence of the 
form factors, we will use the on-shell values, i.e. the pole residua 
$A_i^{(v)} \equiv A_i^{(v)} (m_v^2)$, which are listed in Table I. 
(The singularity of $L(q^2)$ at timelike $q^2$ is an artefact of the 
parametrization without physical significance \cite{mer95}.)

The $\omega$ and $\phi$ couplings $B_i^{(\omega,\phi)}$ can now be 
calculated from the $A_i^{(\omega,\phi)}$ via Eqs. (\ref{omegcoupl}) 
and (\ref{phicoupl}). The new values of the coupling and mixing 
parameters $\kappa g_i$ and $\eta_i$,
\be
\tan(\eta_1 + \epsilon) = 0.4583, \quad \tan(\eta_2 + \epsilon) = 
0.6137, \qquad \kappa g_1 = 5.3218, \quad \kappa g_2 = -0.9294,
\ee
establish the relation between the isocalar and strange current 
couplings. The results for the latter are given in the last 
rows of Table II. Together with the pole masses $m_\omega, m_\phi 
\, \mbox{and} \, m_3$ they form the common parameter set on which all 
the following descriptions of the spectral functions will be based. 

The two remaining couplings $B_{1,2}^{(3)}$ of the 3-pole ansatz are 
again fixed by imposing the asymptotic conditions (\ref{as1}) and 
(\ref{as2}), and their values are also listed in Table II. Note,
incidentally, that logarithmic corrections to the $B_i^{(v)}$ 
(analogous to those for the isoscalar couplings (\ref{momcoup})) 
would not improve the description of the strange form factors 
since the 3-pole ansatz does not even reproduce the asymptotic power 
behavior predicted by QCD (see below). 

We are now ready to discuss the updated results. As the new values for 
the strangeness radius and magnetic moment we find  
\be
r_s^2 = 0.22 \,{\rm fm}^2, \qquad \,\, (r_s^2)_{Sachs} = 0.20 \, 
{\rm fm}^2 , \qquad \,\, \mu_s = -0.26. 
\label{mom}
\ee
Comparison with the values of Section \ref{3pol} shows that the square 
radius increases by 40 \% whereas the absolute value of the magnetic 
moment is reduced by about 20 \%. The bulk of these changes can be traced 
to differences in the values of H{\"o}hler's and MMD's $\phi$-nucleon 
couplings. While both fits find surprisingly large values for these 
couplings\footnote{The large $g_i (\phi N N)$ were interpreted as one 
of the first indications for OZI-rule violations in the nucleon 
\cite{gen76}, although they might at least partially originate from 
$K \bar{K}$ continuum strength \cite{gar92}.},  MMD's $A_1^{(\phi)}$ 
still exceeds those of all H{\"o}hler fits and is almost 30 \% larger 
than their average $(\bar{A}_1^{(\phi)})_H$. The Pauli coupling, on the 
other hand, becomes smaller in the new fits,
\be
\frac{ (A_1^{(\phi)})_{MMD} }{ (\bar{A}_1^{(\phi)})_H} = 1.28, 
\quad \qquad
\frac{ (A_2^{(\phi)})_{MMD} }{ (\bar{A}_2^{(\phi)})_H} = 0.84. 
\ee
The strange current couplings $B_i^{(\phi)}$ move in the same 
directions and bring about most of the mentioned changes in the values 
of the moments.

It would clearly be useful to have a reliable error estimate for the 
updated moments. Unfortunately, however, it is practically impossible 
to asses the systematic errors associated with the input data, the fit 
procedure, the limitations of the 3-pole ansatz etc. Any error analysis 
would therefore necessarily be incomplete and potentially misleading. 
Jaffe's partial error estimate from the variance between different 
isoscalar form factor fits, incidentally, cannot be applied to the 
results (\ref{mom}) since MMD provide only one set of fit parameters. 

Figs. 1a and 1b show the momentum dependence of the updated 3-pole form 
factors at spacelike momenta up to $Q^2 = 1 {\rm GeV}^2$, which includes 
the range to be probed in the Bates, CEBAF and MAMI experiments. For 
comparison these figures also contain the 3-pole form factors of the 
last section with input from H{\"o}hler {\it et al.}'s fits 7.1, 8.1 
and 8.2. 

The new Dirac form factor follows rather closely the one derived from 
H{\"o}hler {\it et al.}s fit 8.2 (which has by far the largest $\phi$ 
coupling under  H{\"o}hler's fits), while it is up to twice as large 
as those from the fits 7.1 and 8.1. The updated Pauli form 
factor, on the other hand, is similar to those from both fits 8.1 and 
8.2 (although the $m_3$ of these fits differ by 400 MeV), whereas it has 
almost twice the size of that from fit 7.1. The average deviation of the 
updated form factors from the ones based on H{\"o}hler's fits is not 
insignificant and leads to the changed predictions (\ref{mom}).

It is instructive to compare the variations under the fits to the 
isocalar {\it electromagnetic} form factors with the induced variations 
between the strange form factors. In the considered momentum range the 
former vary by at most a few percent while the corresponding strange 
form factors show much larger differences. The main source of these 
differences lies in the increased sensitivity of the strange form factor 
analysis to the $\phi$ couplings. 

Also the detailed momentum dependence of the form factors (beyond 
their overall scale) depends sensitively on the relative sizes of the 
pole couplings. The leading $1/q^2$ dependence of the $\phi$ pole, for 
example, cannot be cancelled by the $\omega$ pole alone since the $| 
B_i^{(\omega)} |$ are about an order of magnitude smaller than the $| 
B_i^{(\phi)} |$. In order to satisfy the asymptotic constraints 
(\ref{as1}) and (\ref{as2}), the third--pole couplings must therefore 
have a comparable magnitude, but the opposite sign of the $\phi$ couplings. 
A glance at Table II shows that this is indeed the case: $ B_1^{(3)} 
\simeq - B_1^{(\phi)}$ and $ B_2^{(3)} \simeq - (1/3) B_2^{(\phi)}$. 
One consequently expects a dipole behavior for $F^{(s)}_2$ with a mass 
scale between $m_2$ and $m_3$, and an almost perfect fit for all 
space-like momenta is obtained in the form 
\be
F_1^{(s)} (q^2) &=& \frac16 \frac{r_s^2 q^2}{( 1-\frac{q^2}{M_1^2} )^2},  
\label{dip13p} \\
F_2^{(s)} (q^2) &=& \frac{\mu_s}{( 1-\frac{q^2}{M_2^2} )^2},
\label{dip23p}
\ee
with $M_1 = 1.31 \, {\rm GeV} \simeq M_2 = 1.26 \, {\rm GeV}$ (for the 
MMD parameters, i.e. with $r_s^2= 5.680 \,{\rm GeV}^{-2}$ and $\mu_s = 
-0.257$). The above parametrization also realizes the asymptotic 
behavior (\ref{as1}), (\ref{as2}) explicitly. 

It is well known that an analogous ``bump-dip'' structure, generated by 
the $\omega$ and $\phi$ poles, leads to the approximate dipole behavior of 
the electromagnetic form factors. As a consequence, the dipole mass 
parameter lies between the $\omega$ and $\phi$ masses and the couplings 
to the third pole are about an order of magnitude smaller (cf. Table I). 
The strange form factors are thus much harder, and this probably rather 
generic feature should be kept in mind if one chooses to parametrize 
them in dipole form\footnote{Sometimes Galster dipole parametrizations 
\cite{gal71} with the same mass as in the electromagnetic form factors 
have been used in the literature, e.g. in \cite{gar93}.}. 

We will argue in the next section that the 3-pole results probably 
overestimate the strange form factors, because constraints from QCD 
demand a faster asymptotic decay than (\ref{as1}), (\ref{as2}). Since 
the slow falloff also casts doubt on the results at small momenta (see 
below) it would be desirable to realize at least the maximal decay 
power of the 3-pole ansatz by imposing the additional superconvergence 
conditions $\sum B_1^{(v)} m_v^2 = 0$ and $\sum B_2^{(v)} m_v^4 = 0$. 
However, no free parameters are left to accomodate further constraints, 
and Table I shows that the given parameters do not conspire to satisfy 
them automatically: $\sum B_1^{(v)} m_v^2 = -2.87 \, {\rm GeV}^2$ and 
$\sum B_2^{(v)} m_v^4 = 0.72 \,{\rm GeV}^4$.

\section{Beyond the 3-pole ansatz}
\label{b3pol}

In the remainder of this paper we will discuss minimal extensions of 
the 3-pole ansatz which correct the asymptotic $q^{2}$ dependence of 
the form factors in order to match the power behavior predicted by 
QCD counting rules \cite{lep80}. 

Let us start with a general comment. Even if the 3-pole description of 
the spectral functions is incomplete since it cannot reproduce the QCD 
asymptotics, this need not necessarily imply an inadequate description 
of the form factors in the {\it low-momentum region} ($Q^2 \le 1  {\rm 
GeV^2}$) probed at Bates, MAMI and CEBAF. Indeed, the impact of the 
correct asymptotics on the behavior at low momenta is not {\it a priori} 
clear: it depends both on how much of this behavior is determined by 
just the low-$s$ strength in the spectral functions and on how 
exclusively the asymptotic decay originates from higher-lying strength. 
These two questions are of course directly related to the (at present 
unknown) momentum scale at which the asymptotic behavior sets in. 

Moreover, even the question of {\it which} asymptotic behavior to 
implement for an optimal low-momentum description of the form factors 
depends on this scale. QCD dimensional counting rules 
\cite{bro75,lep80} predict that elastic interactions of the strangeness 
current at large spacelike $q^2$ are (up to logarithms) suppressed as 
$(1/q^2)^n$, where $n$ corresponds to the number of hard gluon propagators 
needed to distribute the momentum transfer of the probe in the target 
nucleon. At very large $Q^2= - q^2$ the leading asymptotic power behavior 
arises from extrinsic\footnote{For an early discussion of intrinsic and 
extrinsic contributions in the context of charm quark admixtures see 
Ref. \cite{bro81}. I am indebted to Stan Brodsky for very helpful 
discussions on this subject.} radiative corrections, which renormalize 
the strangeness current. Thus they are suppressed by additional factors 
of the strong coupling $\alpha_s$ and decay with the same powers of 
$(1/q^2)$ as the electromagnetic form factors, i.e.
\be
F_1^{(s)} (q^2) \rightarrow \left( \frac{-1}{q^2}\right)^2, \qquad \quad
F_2^{(s)} (q^2) \rightarrow \left( \frac{-1}{q^2}\right)^3. 
\label{asymex} 
\ee

However, enforcing this behavior might not be the best choice for an 
optimal description of the strange form factors at {\it small and 
intermediate} momenta in the pole approximation. Alternatively, one could 
implement the large-$Q^2$ behavior of the intrinsic contributions, which 
originate from $s \bar{s}$ admixtures to the nucleon wave function. 
Although the intrinsic contributions are asymptotically subleading,
\be
F_1^{(s)} (q^2) \rightarrow \left( \frac{-1}{q^2}\right)^4, \qquad \quad
F_2^{(s)} (q^2) \rightarrow \left( \frac{-1}{q^2}\right)^5,
\label{asymin}
\ee 
they lack the additional $\alpha_s$--suppression of their extrinsic 
counterparts and might thus determine the momentum dependence of the form 
factors at intermediate spacelike $q^2$, i.e. before the extrinsic 
behavior ultimately begins to dominate. In this situation, the form 
factors would up to the onset of the extrinsic behavior be 
better described by imposing the intrinsic asymptotics (\ref{asymin}). 
The mismatch with the extrinsic behavior at very large momenta, 
furthermore, would then have very little impact in the low--momentum 
region of interest. 

An early transition to the asymptotic regime improves both the accuracy 
of the pole approximation and the chances for the appearence of an 
intrinsic region at intermediate momenta. At present, however, it is 
unknown at which momenta the asymptotic behavior sets in and whether 
intrinsic or extrinsic contributions dominate initially. In the 
following sections we will therefore consider both possibilities and 
analyze the minimal extensions of the 3-pole ansatz which implement 
either of the asymptotics (\ref{asymex}) or (\ref{asymin}). 

In the pole approximation (\ref{npsd}) these extensions correspond to 
the implementation of additional poles, which represent both 
higher--lying resonances and strength from multi-particle intermediate 
states like $(2n+1) \pi, K \bar{K}, N \bar{N}$ and $\Lambda 
\bar{\Lambda}$. An explicit inclusion of continuum cuts (in particular 
from the $K \bar{K}$ states) is beyond the scope of the present paper 
and will be reserved for a separate investigation \cite{for96}. 

\subsection{The 4-pole ansatz with extrinsic asymptotics}
\label{4pol}

The phenomenological values of the three lowest--lying masses and of the 
$\phi$ and $\omega$ couplings (as found in Section \ref{3polup}) imply 
that at least four poles are required to realize the asymptotic behavior 
(\ref{asymex}). The minimal ansatz for the form factors with extrinsic 
asymptotics is therefore 
\be
F_i^{(s)} (q^2) = \sum_{v=1}^4 \, B_i^{(v)} \, \frac{m^2_v}{m^2_v - q^2},
\qquad i \in \{1,2\},
\label{f4p}
\ee
together with the two superconvergence relations 
\be
\sum_{v=1}^4 B_2^{(v)} m_v^{2n} = 0,  \qquad \qquad n = \{ 1,2 \} 
\label{sconv2}
\ee
for the (unsubtracted) Pauli form factor and the normalization 
and superconvergence conditions
\be
\sum_{v=1}^4 B_1^{(v)} m_v^{2n} = 0,  \qquad \qquad n = \{0,1 \}
\label{sconv4p}
\ee
for the Dirac form factor. 

Since the couplings $B_i^{(\omega,\phi)}$ and the first three masses 
are already fixed, there are five parameters (four couplings and one 
mass) left to be determined. Writing the above constraints in the 
form
\be 
\left(\begin{array}{cc} m_3^2 & m_4^2 \\ m_3^4 & m_4^4 \end{array} 
\right)
\left(\begin{array}{cc}  m_3^{-2} & 0  \\ 0 & m_4^{-2} \end{array} 
\right) 
\left(\begin{array}{c} B_1^{(3)} \\ B_1^{(4)} \end{array} \right) = 
- \left(\begin{array}{c} C_1^{(3)} \\ C_1^{(4)} \end{array} \right) 
\label{scr1matrix}
\ee
and
\be 
\left(\begin{array}{cc} m_3^2 & m_4^2 \\ m_3^4 & m_4^4 \end{array} 
\right) 
\left(\begin{array}{c} B_2^{(3)} \\ B_2^{(4)} \end{array} \right) = 
- \left(\begin{array}{c} C_2^{(3)} \\ C_2^{(4)} \end{array} \right)  
\label{scr2matrix}
\ee
($C_1^{(i)} \equiv \sum_{j=1}^2 B_1^{(j)} m_j^{2(i-3)}$, $C_2^{(i)} 
\equiv \sum_{j=1}^2 B_2^{(j)} m_j^{2(i-2)}$) shows immediately that 
these equations have a unique solution for the couplings $B_i^{(3,4)}$ 
as a function of the mass $m_4$. (The Vandermonde mass matrices are 
regular since the 4-pole ansatz implies $m_3 \neq m_4 $.) 
This solution is
\be
B_1^{(v)} &=& - B_1^{(1)} \, \prod_{i \neq v} 
\frac{m_1^2-m_i^2}{m_v^2-m_i^2} - B_1^{(2)} \, 
\prod_{i \neq v} \frac{m_2^2-m_i^2}{m_v^2-m_i^2} ,  
\label{4pcoupl1} \\
B_2^{(v)} &=& - B_2^{(1)}  \,\frac{m_1^2}{m_v^2}\, 
\prod_{i \neq v} \frac{m_1^2-m_i^2}{m_v^2-m_i^2} - B_2^{(2)}  
\,\frac{m_2^2}{m_v^2}\, \prod_{i \neq v} 
\frac{m_2^2-m_i^2}{m_v^2-m_i^2} . 
\label{4pcoupl2}
\ee
($v,i \in \{3,4\}$.) 

After eliminating the $B_i^{(3,4)}$ from the constrained 4-pole ansatz 
with the help of (\ref{4pcoupl1}) and (\ref{4pcoupl2}), one obtains 
the form factors as functions of the already determined masses and 
couplings, as well as of $m_4$:
\be
F_1^s(q^2) = q^2 \left[\frac{B_1^{(1)} }{m_1^2-q^2} \prod_{i=3}^4 
\frac{m_i^2-m_1^2}{m_i^2-q^2} +  \frac{B_1^{(2)}}{m_2^2-q^2} 
\prod_{i=3}^4 \frac{m_i^2-m_2^2}{m_i^2-q^2} \right], \label{4pf1}
\ee 
\be
F_2^s(q^2) =\frac{B_1^{(1)} m_1^2}{m_1^2-q^2} \prod_{i=3}^4 
\frac{m_i^2-m_1^2}{m_i^2-q^2} +  \frac{B_1^{(2)} m_2^2}{m_2^2-q^2} 
\prod_{i=3}^4 \frac{m_i^2-m_2^2}{m_i^2-q^2} .\label{4pf2}
\ee
These expressions for the strange form factors with extrinsic 
asymptotics are useful since they contain only unconstrained 
parameters. 

Before turning to their numerical evaluation we still have to determine 
the position of the fourth pole. In contrast to the three low--lying 
poles, theoretical estimates of its mass would have to rely on 
uncontrolled assumptions since the 3--pole fits to the electromagnetic 
form factors contain no information about this pole. Relating the 
couplings of the third and fourth pole in a mixing model (e.g. for the 
$\omega(1600)$ and the $\phi(1680)$) in order to determine $m_4$ from 
Eqs. (\ref{4pcoupl1}) or (\ref{4pcoupl2}), for example, would be 
unreliable since these resonances have a large width and since their 
flavor content can be estimated only roughly. 

Alternatively, one could attempt to fix the fourth pole mass at the 
value of another known ($0^- 1^{- -}$) resonance. Although it becomes 
increasingly unreliable for higher--lying poles, a similar procedure 
was followed in Ref. \cite{dub92} to estimate the electromagnetic form 
factors. 

In the strangeness channel, however, not all of the five established 
resonances seem to be good candidates for additional poles. The 
$\omega(1420)$ did not require a pole in the MMD fit\footnote{Only the 
third pole of H{\"o}hler {\it et al.}s Fit 8.1 at 1.4 GeV gives some 
indication for a (small) contribution from the $\omega(1420)$.} and thus 
appears to couple only weakly to the nucleon. Due to its dominant flavor 
content (as estimated from the observed partial decay widths), moreover, 
it will probably couple even weaker to the strangeness current. The 
remaining two established $0^- 1^{--}$ resonances, the $\omega(1600)$ and 
the $\phi(1680)$, cannot be resolved in the pole approximation since 
their widths of 280 and 150 MeV are much larger than their mass 
difference. Thus only three strong and well separated centers of strength 
are known in the isoscalar channel. Since those are already represented 
by the first three poles, they offer no guidance in finding the value of 
$m_4$. 

For the following numerical estimates we will therefore restrict the 
value of $m_4$ only by the requirement that it should exceed the third 
pole mass ($m_3 = 1.6 \,{\rm GeV}$) by at least a typical 
width\footnote{Continuum strength around these invariant masses would 
probably have an even broader distribution.} of $ \sim 300 \, {\rm MeV}$. 
Under this condition neighboring strength positions do not coalesce and 
have to be described by separate poles. For masses $1.9 \, {\rm GeV} \le 
m_4 \le \infty$ we find the strangeness radius and magnetic moment to 
interpolate monotonically in the ranges 
\be 
0.15 \, {\rm fm}^2 \le &r_s^2& \le 0.22 \, {\rm fm}^2, \nonumber \\  
0.14 \, {\rm fm}^2 \le &\,\, (r_s^2)_{Sachs}\,\,& \le 0.20 \, 
{\rm fm}^2,  \nonumber \\
-0.18 \ge &\mu_s& \ge -0.26.
\label{4pbounds}
\ee
If $m_4$ becomes large, the fourth pole does not affect the momentum 
dependence of the form factors at $Q^2 \ll m_4^2$. For $m_4 \rightarrow 
\infty$ the 4--pole form factors become identical to those from the 
3--pole ansatz of Section \ref{3polup}, which thus provide the upper 
bounds on $r_s^2$ and $| \mu_s |$ in (\ref{4pbounds}). (To the quoted 
accuracy these values are reached at $m_4 \sim 12 \,{\rm GeV}$.) A fourth 
pole in the region around $2 \,{\rm GeV}$, on the other hand, reduces 
the 3-pole predictions for $r_s^2$ and $| \mu_s |$ by about one 
third\footnote{Even if one chooses $m_4$ overly close to $m_3$ in order 
to minimize $r_s^2$ and $| \mu_s |$, their values cannot be reduced much 
further. For $m_4 = 1.65 \,{\rm GeV}$, e.g., one finds  $r_s^2 = 0.12 \, 
{\rm fm}^2$ and $\mu_s = -0.15$.} while the signs of the form factors 
and moments remain unchanged. 

Due to the asymptotic constraints on the 4--pole spectral densities, 
the size of the third--pole couplings remains comparable to that of the 
$\phi$ couplings for all admissable values of $m_4$ (cf. Eqs. 
(\ref{4pcoupl1}), (\ref{4pcoupl2})). In view of MMD's identification of 
this pole with the $\omega(1600)$ it is surprising that these couplings 
are much larger than those of the $\omega(782)$ (even if growing 
continuum contributions to higher--lying poles should be expected). 
This suggests that in the strangeness channel the third pole receives 
significant strength from the $\phi(1680)$ resonance, which presumably 
has much stronger couplings to the strange current. 

If $m_4$ lies in the region around $ 2 \,{\rm GeV}$ even the fourth--pole 
couplings reach a size comparable to those of the second and third pole. 
In this case one expects the form factors to show a quadrupole momentum 
dependence. For the conservative choice $m_4=1.9 \, {\rm GeV}$ (i.e. 
for the (in absolute magnitude) smallest 4-pole form factors with 
extrinsic asymptotics) we indeed find to a very good approximation 
\be
F_1^{(s)} (q^2) &=& \frac16 \frac{r_s^2 q^2}{( 1-\frac{q^2}{M_1^2} )^3} ,  
\label{dip14p} \\
F_2^{(s)} (q^2) &=& \frac{\mu_s}{( 1-\frac{q^2}{M_s^2} )^3},
\label{dip24p}
\ee
with $r_s^2 = 0.1482 \,{\rm fm}^2$, $\mu_s = -0.1789$,  $M_1 = 1.47 
\,{\rm GeV}$ and $M_2 = 1.43 \,{\rm GeV}$. The values of the mass 
parameters correspond rather closely to the weighted mean of the three 
dominant pole positions. 

The 4-pole form factors with $m_4=1.9 \,{\rm GeV}$ are shown in 
Figs. 2a and 2b. For comparison we also show the 3-pole form factors, 
i.e. the $m_4 \rightarrow \infty$ limit of the 4-pole ansatz. All 4-pole 
form factors with $m_4 \ge 1.9 \, {\rm GeV}$ lie in the band between 
these two bounds and are monotonically reached by increasing $m_4 $. 
At $Q^2 \sim 1 \,{\rm GeV}^2$ the 4-pole Dirac form factor is about two 
times smaller than that of the 3-pole ansatz; the strangeness radii 
differ by a factor of 0.7. 

From the figures it is apparent that the weaker (and incorrect) 
asymptotic decay of the 3-pole bounds affects the momentum dependence of 
the form factors already at small and intermediate momenta. Thus the 
3--pole predictions probably overestimate the strange form factors and 
the magnitude of their moments. 

\subsection{The 6-pole ansatz with intrinsic asymptotics}
\label{6pol}

Realizing the intrinsic decay (\ref{asymin}) by an ansatz of the form 
(\ref{npsd}) requires minimally five poles. The 5-pole ansatz, however,
(which has the attractive feature that the number of its free parameters 
is matched by an equal number of constraints) turns out to be 
overconstrained: no exact solutions of the constraints exist 
and the approximate ones have unphysical features. (In particular, they 
require a pronounced additional pole in close vicinity to the $\phi$, 
but with couplings of opposite sign. A detailed discussion of the 
5-pole ansatz is relegated to appendix \ref{full5pole}.)

Therefore, the minimal description of the intrinsic asymptotics in 
the pole approximation is given by the 6-pole ansatz
\be
F_i^{(s)}(q^2) = \sum_{v=1}^6 \, B_i^{(v)} \, \frac{m^2_v}{m^2_v - q^2},
\qquad i \in \{1,2\}
\label{f6p}
\ee
with the constraints 
\be
\sum_{v=1}^6 B_1^{(v)} m_v^{2k} &=& 0,  \qquad k = \{0,1,2,3 \}, \\
\sum_{v=1}^6 B_2^{(v)} m_v^{2n} &=& 0,  \qquad  n = \{ 1,2,3,4 \}.
\label{sconv6p}
\ee
We solve these constraints by a straighforward extension of the approach
described in Section \ref{4pol}. The resulting expressions for the 
couplings and form factors are generalizations of (\ref{4pcoupl1}), 
(\ref{4pcoupl2}) and (\ref{4pf1}), (\ref{4pf2}), with the product indices 
now ranging up to $N=6$. The couplings are again uniquely determined as 
functions of the masses, but two more pole positions, $m_5$ and $m_6$, 
are left a priori unrestricted.

As in the last section, we base the further analysis of the 6-pole 
form factors on the premise that higher-lying poles should have a minimal 
mass difference of $\Delta m = 300 \, {\rm MeV}$. Under this assumption 
the most conservative form factor estimate (i.e. the one smallest in 
magnitude) corresponds to the mass values $\{m_4, m_5, m_6 \} 
= \{1.9, 2.2, 2.5 \} \,{\rm GeV}$. Increasing $m_6, m_5$ and $m_4$ up to 
infinity leads to monotonically increasing magnitudes of the form factors 
(and thus of $r_s^2$ and $| \mu_s |$) until finally, via the 5-pole and 
4-pole results, the 3-pole form factors are reached as upper bounds. In 
the considered domain of pole positions we find the leading moments in 
the ranges 
\be 
0.089 \, {\rm fm}^2 \le &r_s^2& \le 0.22 \, {\rm fm}^2, \nonumber \\  
0.081 \, {\rm fm}^2 \le & \,\,(r_s^2)_{Sachs} \,\,& \le 0.20 \, 
{\rm fm}^2, \nonumber \\  -0.114 \ge &\mu_s& \ge -0.26.
\label{6pbounds}
\ee
Relative to the 3-pole estimates the intrinsic asymptotics can thus 
reduce the size of the strangeness radius and magnetic moment by up to 
a factor of 2.5, i.e. considerably more than the extrinsic asymptotics. 

As in the case of the 3-pole and 4-pole ans{\"a}tze before, we find the 
6-pole form factors well fitted by the simplest multipole formulae which 
match their asymptotic behavior: 
\be
F_1^{(s)} (q^2) &=& \frac16 \frac{r_s^2 q^2}{( 1-\frac{q^2}{M_1^2} )^5} ,  
\label{5p1} \\
F_2^{(s)} (q^2) &=& \frac{\mu_s}{( 1-\frac{q^2}{M_s^2} )^5},
\label{5p2}
\ee
with $r_s^2 = 0.08879 \,{\rm fm}^2$ and $\mu_s = -0.1136$. For $\{m_4, 
m_5, m_6 \} = \{1.9, 2.2, 2.5 \} \,{\rm GeV}$, in particular, we find 
$M_1 = 1.72 \,{\rm GeV}$ (for $Q^2 \le 10 \,{\rm GeV}$ a somewhat 
better fit 
is obtained with $M_1 = 1.79 \,{\rm GeV}$) and $M_2 = 1.66 \,{\rm GeV}$. 
Similarly to the previously discussed cases, the momentum dependence of 
Eqs. (\ref{5p1}) and (\ref{5p2}) reflects the fact that all five 
couplings $B_i^{(2)} - B_i^{(6)}$ are of comparable size and have 
alternating signs.

Figures 2a and 2b contain, in addition to the 3- and 4-pole form factors,
the exact 6-pole form factors with the above set of higher-lying strength 
positions, $\{m_4, m_5, m_6 \} = \{1.9, 2.2, 2.5 \} \,{\rm GeV}$. The 
graphs demonstrate the discussed hierarchy of magnitudes. The bands 
between the lower bounds from the 6-pole form factors and the upper 
bounds from the 3-pole form factors can be continuously covered by 
increasing the three highest-lying masses of the 6-pole ansatz from 
their minimal values to infinity. 

\section{Generic aspects of the leading moments}
\label{gen}

The leading nonvanishing moments characterize the low--momentum behavior
of the strange vector form factors and set their scale. They will also 
be the first to be measured in the forthcoming experiments, and for both 
reasons they are the focus of most current theoretical work in the 
strangeness vector channel. 

Despite these efforts, however, no consensus on the size or even on 
the signs of these moments has been reached. Nucleon model predictions, 
in particular, involve large theoretical uncertainties and led to both 
positive and negative signs for the strangeness radius\footnote{The 
mentioned model calculations find, with the exception of Ref. \cite{hon93}, 
a negative sign for the strange magnetic moment. Hadron models do agree 
better on the value of $\mu_s$ than on that of $r_s^2$, probably because 
the moment is less sensitive to details of the strangeness distribution.}. 
Model--independent information on the sign of the moments would thus be 
very useful. Moreover, it could give valuable hints towards the dynamical 
origin of the nucleon's strangeness distribution. 

A kaon cloud (e.g. a $K-\Lambda$ component in the nucleon wave function 
\cite{mus93,nos,for94}), for example, generates a negative contribution 
to the radius. Neglecting recoil effects, this can be qualitatively 
understood from the fact that the kaon is less than half as heavy 
as the $\Lambda$ (or any other hyperon). Its strangeness distribution 
thus reaches out farther from the nucleon's center of mass, and its 
(in our convention) negative strangeness charge determines the sign of 
$r_s^2$. An analogous argument (with the even lighter pions in the 
cloud) has been used to explain the negative charge radius of the 
neutron. Garvey {\it et al.}'s reanalysis of the BNL neutrino scattering 
data \cite{gar93}, incidentally, seems to favor a negative sign of the 
strangeness radius, although the statistics of the data is too poor to 
reach a reliable conclusion. The majority of the present nucleon model 
calculations also find a negative $r_s^2$.

All our dispersive estimates, on the other hand, led (independently of 
the number of poles involved and of the required asymptotic behavior) to 
the opposite, positive sign of the strangeness radius. This (as well as 
the negative sign of $\mu_s$) is a rather generic property of the pole 
approximation, which can be read off from the general N-pole expressions 
for the lowest moments, 
\be
r_s^2 = \sum_v^N \frac{ B_1^{(v)} }{m_v^2}, \qquad \qquad \mu_s =
 \sum_v  B_2^{(v)}. \label{momexpl}
\ee
These expressions imply that the signs of the characteristically large 
$\phi$ couplings (which are positive (negative) in the Dirac (Pauli) 
form factor) determine the signs of the leading moments. The impact of 
the higher pole couplings is attenuated by their alternating sign (as 
required by the superconvergence relations) and by the mass factor in 
the expression for $r_s^2$. This explains also why we find the signs of 
the moments to be independent of the number of higher-lying poles.

The opposite signs of the pole and kaon--cloud predictions for $r_s^2$
might point towards missing physics in either framework. A positive 
sign from experiment could indicate, for example, that the $K-\Lambda$ 
intermediate states -- although they are the lightest accessible ones 
-- do not provide the main contribution to the nucleon's strangeness 
distribution. Indeed, their impact could be overcompensated by the more 
numerous intermediate states with higher mass in which the negative 
strangeness is carried by the heavier particle. This issue is currently 
under investigation \cite{for952}. 

A negative sign of the strangeness radius (due to the kaon cloud 
contribution or any other mechanism), on the other hand, would 
expose a serious shortcoming of the pole approximation, probably due to 
the neglect of continuum contributions. Indications in this direction 
come from the generalized vector meson dominance framework of Ref. 
\cite{for94}, which emphasizes the importance of the $K \bar{K}$ 
continuum in addition to the $\omega$ and $\phi$ poles. In this 
approach the kaonic intermediate states are consistently incorporated 
through extended vector--meson nucleon vertices which describe the 
intrinsic strangeness distribution of the nucleon. Despite the large 
and positive Dirac coupling of the $\phi$ these intrinsic contributions 
turn the sign of $r_s^2$ negative. This finding suggests that the role 
of $K \bar{K}$ continuum states in the dispersive analysis should be 
examined explicitly \cite{for96}. 

\section{Summary and Conclusions}
\label{sum}

We have analyzed the nucleon's strange vector form factors in a 
dispersive approach which circumvents dynamical model calculations and 
relies instead on experimental input from isoscalar electromagnetic form 
factor data. We emphasize in particular the impact of QCD-imposed 
constraints on the asymptotic behavior of the form factors. 

All intermediate states through which the strangeness current couples 
to the nucleon, including higher-lying resonance and continuum 
contributions, are described in the pole approximation, i.e. 
by isocalar vector meson states of zero width. This amounts to a 
generalization of the vector meson dominance principle which 
successfully accounts for electromagnetic interactions of hadrons.
Because of its largely generic character we expect this approach to be 
a useful starting point also for estimates of the strange form factors.  

After updating the results of Jaffe's minimal 3--pole ansatz with 
input from a new fit to the world data set on the electromagnetic 
form factors, we extend the pole approximation in order to implement 
the asymptotic momentum dependence which QCD counting rules predict. 
In the following we collect and discuss our main results and comment 
on some directions for their refinement: 

\begin{enumerate}
\renewcommand{\labelenumi}{\arabic{enumi})}

\item
The update of Jaffe's 3-pole analysis on the basis of the current world 
data set for the electromagnetic form factors leads to a by 40 \% 
increased strangeness radius and a by 20 \% reduced magnitude of the 
strange magnetic moment:
\be
r_s^2 = 0.22 \,{\rm fm}^2, \qquad (r_s^2)_{Sachs} = 0.20 \,{\rm fm}^2,
\qquad \mu_s = -0.26.  \nonumber
\ee

The considerably extended data base of the new fits and their more 
complete description of the isoscalar form factor asymptotics should 
improve the accuracy of these results.

Since both the $\phi$ pole and the third pole contribute with comparable 
weight and with opposite sign to the spectral functions, the 3-pole form 
factors have, to a very good approximation, dipole form. The dipole mass 
parameter lies between the masses of the two dominant poles at about 
$1.3 \,{\rm GeV}$, which makes the strange form factors considerably 
harder than the electromagnetic ones (with a cutoff mass of $0.84 
\,{\rm GeV}$). This presumably rather generic property of the strange 
form factors should be kept in mind if one chooses to parametrize their 
momentum dependence in dipole form. 

\item
The main advantage of the 3-pole ansatz lies in its simplicity. On the 
other hand, it cannot describe the fast decay of the strange form 
factors which QCD counting rules predict at asymptotically large momenta. 
In order to correct this shortcoming one has to implement either the 
asymptotic behavior of the ultimately dominating extrinsic contributions 
or the faster decay of the intrinsic ones, which may determine the 
momentum dependence at intermediate momenta. We consider both 
alternatives since it is at present unclear which strategy will lead to 
a better description at the low and intermediate momenta of interest. 

The asymptotics of the extrinsic contributions requires minimally four 
poles, and with the conservative choice $m_4  = 1.9 \,{\rm GeV}$ for 
the fourth pole position one finds
\be 
r_s^2 = 0.15 \, {\rm fm}^2, \qquad  
(r_s^2)_{Sachs} = 0.14 \, {\rm fm}^2, \qquad 
\mu_s = -0.18 , \nonumber
\ee
which amounts to a reduction of the 3-pole results by about a third. 
Increasing the value of $m_4$ leads to monotonically increasing values 
for $r_s^2$ and $| \mu_s|$ and, in the limit $m_4 \rightarrow \infty$, 
back to the 3-pole results. 

Minimally six poles are needed in order to realize the intrinsic 
asymptotics, and for a similarly conservative choice $\{m_4, m_5, m_6 
\} = \{1.9, 2.2, 2.5 \} \,{\rm GeV}$ of the higher pole positions 
one obtains 
\be 
r_s^2 = 0.089 \, {\rm fm}^2, \qquad 
(r_s^2)_{Sachs} = 0.081 \, {\rm fm}^2, \qquad 
\mu_s = -0.114. \nonumber
\ee
Compared to the 3-pole results the magnitude of the moments decreases 
by about a factor 2.5, i.e. considerably more than in the case of the 
extrinsic asymptotics. Again the 3-pole results can be recovered by 
shifting the higher-lying pole masses to infinity. 

The above results mark out the range of values for the leading moments 
which are accessible in the pole approximation. The predictions for 
$r_s^2$ are more sensitive to the momentum dependence of the form 
factor and thus probably less reliabe than those for $\mu_s$. The 
momentum dependence of the 3--, 4-- and 6--pole form factors is well 
fitted by the simplest multipole parametrizations which match their 
asymptotic behavior.

\item
For several reasons it does not seem useful to extend the pole 
approximation beyond the minimal 4(6)--pole ans{\"a}tze considered 
above. While each new pole introduces three additional and a priori 
undetermined parameters, there are no further asymptotic constraints 
which could be used to fix them. Moreover, the zero--width approximation 
becomes less reliable for higher--lying strength and the impact of 
additional poles on the low--momentum behavior of the form factors would 
decrease since no further superconvergence conditions keep their 
couplings large. 

\item
The 3-pole ansatz yields the (in absolute magnitude) largest possible 
strange form factors in the pole approximation. Due to the unrealistic 
asymptotics, however, it probably overestimates the size of their 
leading moments by up to a factor 2.5. It is quite remarkable that the 
QCD--prescribed modifications of the asymptotic behavior can have such 
a strong impact on the low--momentum predictions. 

\item
In all the considered pole--ansatz variants the couplings to the third 
pole turn out to be an order of magnitude larger than the corresponding 
couplings in the isoscalar electromagnetic form factors. The reason is 
that the third pole has to balance the strong $\phi(1019)$ pole in order 
to achive even the modest asymptotic decay of the 3--pole form factors
(where no superconvergence constraint is imposed on the normalized Dirac 
form factor). The third--pole couplings are also four to eight times 
larger than the strange couplings to the 
$\omega(781)$. It seems therefore unlikely that the third pole can be 
identified (as in the electromagnetic case) with the $\omega(1600)$, 
which has a mostly nonstrange flavor content. An alternative and more 
convincing provenance for at least part of this strength could be the 
neighboring $\phi(1680)$. (Of course, both resonances cannot be 
distinguished in a pole ansatz since their widths are much larger than 
their mass difference.) If correct, this hypothesis would imply that 
the two established $\phi$ resonances alone determine to a large part 
the low-momentum behavior of the strange vector form factors.

\item
The signs of the leading form--factor moments turn out to be independent 
of the number of poles and of the implemented asymptotics: in the pole 
approximation they follow those of the large $\phi$ couplings and yield 
a {\it positive} square strangeness radius and a negative strange magnetic 
moment. These result underline once more the crucial importance of the 
$\phi(1019)$ resonance for the behavior of the vector form factors. 

Most model calculations agree with the negative sign of the strange 
magnetic moment (see Table III), but the majority of them predicts the 
opposite, negative sign for the strangeness radius. This might point 
towards the relevance of the so far neglected $K \bar{K}$ intermediate 
states in the dispersive analysis. On the basis of its large strangeness 
content and comparatively small 
invariant mass, the $K \bar{K}$ continuum can in any case be expected to 
generate a significant low-energy cut. Its inclusion into the dispersion 
analysis is in progress \cite{for96}. Other relevant contributions to the 
spectral densities could come from the $(2n+1) \pi, N \bar{N}, \Lambda 
\bar{\Lambda}$ (and possibly higher--lying) continuum cuts. An explicit 
estimate of these contributions could test if such cuts are sufficiently 
well accounted for by zero-width poles, as we have assumed. 

\end{enumerate}

Our main intent in the present study was to examine the impact of the 
strange form factor asymptotics on the low-momentum predictions of the 
dispersive analysis. In particular, we determined the range of values 
for the leading moments (in pole approximation) which are consistent 
with our present knowledge of the QCD asymptotics. Even if this range 
remains rather large, some further and more general conclusions can be 
drawn from our analysis. The most important ones are probably (i) that 
the third pole plays (independently of the required asymptotics) a much 
more important role than in the isoscalar electromagnetic form factors, 
(ii) that its interplay with the $\phi(1019)$ makes the strange form 
factors harder than the electromagnetic ones and (iii) that the signs 
of the leading moments, which originate mainly from the $\phi$ pole, 
are generic in the given framework and might point towards limitations 
of the pole approximation. Our findings suggest that the study of 
kaonic continuum states is the most promising direction for further 
extensions of the dispersive strange form factor analysis.

\acknowledgements

I would like to thank Stan Brodsky for helpful correspondence on 
the asymptotic behavior of the strange form factors and Wolfram Weise 
for an interesting discussion. Financial support by the European 
Community through the HCM programme is also gratefully acknowledged. 

\appendix

\section{Rationalized N-pole form factors}
\label{app1}

In this appendix we collect several useful formulae which are 
encountered in deriving the results of the preceding sections. 
We consider a general $N$-pole ansatz
\be
F_i^{(s)} (q^2)=\sum_{v=1}^N \, B_i^{(v)} \, \frac{m^2_v}{m^2_v - q^2}, 
\label{npole}
\ee
for the form factors and impose the normalization and superconvergence 
relations (SCRs)
\be
\sum_{v=1}^N B_i^{(v)} m_v^{2n} = 0,  \quad \quad n = \{0,1,2,...,k \}. 
\label{sconv}
\ee
(Note that the above constraints are homogeneous and do not fix the 
common scale of the couplings $B^{(v)}$.) Of course, the form factors 
(\ref{npole}) can satisfy only a limited number of superconvergence 
relations, which restricts the maximal power of $1/q^2$ in their 
asymptotic decay. For the strange Pauli form factor this power is $N$, 
while the normalization condition at $q^2 =0$ reduces it to $N-1$ for 
the Dirac form factor. 

The leading asymptotic behavior of the Dirac form factor\footnote{A 
similar argument can be given for the Pauli form factor.} can be seen 
more explicitly by writing the normalization condition and the $N - 1$ 
SCRs as a system of $N$ linear equations $M \cdot B = 0$ for the 
coefficient vector $B$. Since $M$ is a $N \times N$ Vandermonde matrix 
with the determinant
\be
det(M) = \prod_{v < w}^{N-1} (m^2_v - m^2_w)
\ee
and since none of the pole masses are equal, the unique solution is 
$B=0$, i.e. the form factor vanishes identically. A nontrivial $F_1$ 
with $F_1 (0)=0$ allows therefore maximally $N-2$ SCRs, and the $N$-pole 
Dirac form factor cannot decay faster than $F_1 \rightarrow 
(1/q^2)^{N-1}$.

In order to exhibit the impact of the normalization and SCR constraints 
on the form factors and their asymptotic behavior explicitly, one can 
rewrite the $N$-pole ansatz (\ref{npole}) in rationalized form. We will 
list below the resulting expressions for the 3-, 4- and 5-pole form 
factors, which can be easily generalized to larger $N$. We have for the 
3-pole ansatz
\be
\sum_{v=1}^3 \, B_i^{(v)} \, \frac{m^2_v}{m^2_v - q^2} &=&
\left( \prod_{v=1}^3 \frac{1}{m^2_v - q^2} \right) 
\Bigg\{ S_i^{(3,0)} \prod_{v=1}^3 m^2_v \,  +  \nonumber \\
&+& (- q^2)  \left[(s^{(3,1)} S_i^{(3,1)} - S_i^{(3,2)} \right] 
 + (- q^2)^2 S_i^{(3,1)}  \Bigg\}, \label{ap3p}
\ee
for the 4-pole ansatz 
\be
\sum_{v=1}^4 \, B_i^{(v)} \, \frac{m^2_v}{m^2_v - q^2} &=&
\left( \prod_{v=1}^4 \frac{1}{m^2_v - q^2} \right) 
\Bigg\{ S_i^{(4,0)}  \prod_{v=1}^4 m^2_v \,  +  \nonumber \\
&+& (- q^2) \left[ \frac12 \left[(s^{(4,1)})^2 - s^{(4,2)} \right] 
S_i^{(4,1)} - s^{(4,1)} S_i^{(4,2)} +  S_i^{(4,3)} \right] \nonumber \\
&+&  (- q^2)^2 \left[ s^{(4,1)}  S_i^{(4,1)} - S_i^{(4,2)} \right]  
+ (- q^2)^3  S_i^{(4,1)} \Bigg\}, \label{ap4p}
\ee
and for the 5-pole ansatz
\be
\sum_{v=1}^5 \, B_i^{(v)} \, \frac{m^2_v}{m^2_v - q^2} 
&=& \left( \prod_{v=1}^5 \frac{1}{m^2_v - q^2} \right) 
\Bigg\{ S_i^{(5,0)} \,  \prod_{v=1}^5 m^2_v +  \nonumber \\
&+& (- q^2) \left[ 
\frac16 S_i^{(5,1)} \left[ (s^{(5,1)})^3 - 3 s^{(5,2)}  s^{(5,1)} + 
2 s^{(5,3)} \right] \right. \nonumber \\
&-& \left. \frac12 S_i^{(5,2)} \left[ (s^{(5,1)})^2 - s^{(5,2)}  \right] 
+  S_i^{(5,3)} s^{(5,1)} 
-  S_i^{(5,4)} \right]   \nonumber\\
&+& (- q^2)^2 \left[ 
\frac12 S_i^{(5,1)} \left[ (s^{(5,1)})^2 - s^{(5,2)}  \right] 
-  S_i^{(5,2)} s^{(5,1)} 
+  S_i^{(5,3)} \right] \nonumber \\
&+& (- q^2)^3 \left[ 
+  S_i^{(5,1)} s^{(5,1)} 
-  S_i^{(5,2)} \right] 
+ (- q^2)^4 S_i^{(5,1)} \Bigg\}, \label{ap5p}
\ee
where we have introduced the abbreviations 
\be
S_i^{(N,n)} = \sum_{v=1}^N \, B_i^{(v)} m^{2 n}_v, \qquad
s^{(N,n)} =  \sum_{v=1}^N \, m^{2 n}_v.
\ee

Note that these expressions for the $S_i^{(N,n)}$ appear in the 
superconvergence relations, so that their impact on the momentum 
dependence of the constrained form factors can be directly red off 
from Eqs. (\ref{ap3p}) -- (\ref{ap5p}). The correctly normalized 4-pole 
form factors with extrinsic asymptotics, for example, satisfy the 
constraints (\ref{sconv4p}) and (\ref{sconv2}) or in our present 
notation
\be
S_1^{(4,m)}  = 0,  \quad m = \{0,1 \}; \qquad \qquad S_2^{(4,n)}  
= 0,  \quad n = \{ 1,2 \},
\label{c1}
\ee
and thus have the rationalized form 
\be
F^{(s)}_1 (q^2) &=& \left( \prod_{v=1}^4 \frac{1}{m^2_v - q^2} \right) 
\Bigg\{ (- q^2) \left[ - s^{(4,1)} S_1^{(4,2)} +  S_1^{(4,3)} \right] 
- (- q^2)^2 S_1^{(4,2)} \Bigg\}, \nonumber \\ 
F^{(s)}_2 (q^2) &=& \left( \prod_{v=1}^4 \frac{1}{m^2_v - q^2} \right) 
\Bigg\{ S_2^{(4,0)}  \prod_{v=1}^4 m^2_v + (- q^2) S_2^{(4,3)} \Bigg\}.
\ee
In this form the cancellation of the asymptotic behavior from 
individual poles up to the required order and the large--$q^2$ 
behavior (\ref{asymex}) become explicit. The rationalized expressions 
(\ref{ap3p}) -- (\ref{ap5p}) confirm the above result for the maximal 
leading decay power of the $N$--pole ansatz. 

\section{The 5-pole ansatz with intrinsic QCD asymptotics}
\label{full5pole}

At least five poles are needed to realize the intrinsic 
asymptotic behavior (\ref{asymin}) in the pole approximation. Besides 
being in this sense minimal, the 5-pole ansatz has the additional 
advantage that the number of normalization and superconvergence 
constraints matches the number of free parameters. A physically 
acceptable solution of these constraints would therefore determine 
the form factors without further input. 

In order to provide the 5-pole ansatz 
\be
F_i^s(q^2) = \sum_{v=1}^5 \, B_i^{(v)} \, \frac{m^2_v}{m^2_v - q^2},
\qquad i \in \{1,2\},
\label{f5p}
\ee
with the correct normalization and the intrinsic asymptotics (i.e. 
$\lim_{Q^2 \rightarrow \infty} Q^{8} F_2^s =0 $ and  $\lim_{Q^2 
\rightarrow \infty} Q^{6} F_1^s =0$), the constraints 
\be
\sum_{v=1}^5 B_2^{(v)} m_v^{2n} = 0,  \quad n = \{ 1 - 4 \}; 
\qquad 
\sum_{v=1}^5 B_1^{(v)} m_v^{2n} = 0,  \quad n = \{0 - 3 \}.
\label{sconv1}
\ee
have to be imposed. Since the 4 couplings $B_i^{(\omega,\phi)}$ and the 
3 masses $m_{\omega}, m_{\phi}, m_3$ are already fixed at the values of 
Section \ref{3polup},  8 parameters are left to be determined from the 8 
constraints (\ref{sconv1}). To this end we first derive the couplings 
$B_i^{(v)}, v \in \{3,4,5\}$ as a function of the masses $m_{3} - m_{5}$ 
by using the 6 constraints with the lowest mass weights and find the 
unique solution 
\be
B_1^{(v)} &=& - B_1^{(1)} \, \prod_{i \neq v} 
\frac{m_1^2-m_i^2}{m_v^2-m_i^2} - B_1^{(2)} \, 
\prod_{i \neq v} \frac{m_2^2-m_i^2}{m_v^2-m_i^2} ,  \label{coupl1} \\
B_2^{(v)} &=& - B_2^{(1)}  \,\frac{m_1^2}{m_v^2}\, 
\prod_{i \neq v} \frac{m_1^2-m_i^2}{m_v^2-m_i^2} - B_2^{(2)}  
 \,\frac{m_2^2}{m_v^2}\, \prod_{i \neq v} 
\frac{m_2^2-m_i^2}{m_v^2-m_i^2}  . \label{coupl2}
\ee
($v,i \in \{3,4,5\}$.) 

The two masses $m_4$ and $m_5$ in the above expressions are related 
by the two remaining constraints, which both have the form of cubic 
equations for $m_4^2$. Out of its 3 real solutions, two are trivial, 
$m_4^2=m_3^2$ and $m_4^2= m_5^2$, and can be discarded. The remaining 
one gives
\be
m_4^2 = m_2^2 \, \frac{1+ \epsilon_1 \, \mu^2 \, M (m_5^2)}{1+ 
\epsilon_1 \, M (m_5^2)}  \qquad {\rm from} \qquad \sum_{v=1}^5 
B_1^{(v)} m_v^6 = 0, \label{m4m51}
\ee
and 
\be
m_4^2 = m_2^2 \, \frac{1+ \epsilon_2 \, \mu^4 \, M (m_5^2)}{1+ 
\epsilon_2 \, \mu^2 \, M (m_5^2)} \qquad {\rm from} \qquad 
\sum_{v=1}^5 B_2^{(v)} m_v^8 = 0,  \label{m4m52}
\ee
where we defined
\be
M(m_5^2) \equiv \frac{(m_1^2- m_3^2)(m_1^2-m_5^2)}{(m_2^2-
m_3^2)(m_2^2-m_5^2)}, \quad \qquad \mu = \frac{m_1}{m_2} = 0.767 .
\ee
  
Either of the above equations (\ref{m4m51}) and (\ref{m4m52}) implies 
characteristic restrictions on the allowed values of $m_4^2$ and $m_5^2$. 
Their origin can be traced to the dominance of the $\phi$ pole over 
the $\omega$ pole, which is reflected in the small coupling ratios
\be
\epsilon_1 \equiv \frac{B_1^{(\omega)}}{B_1^{(\phi)}} = -0.116,
\qquad \epsilon_2  \equiv \frac{B_2^{(\omega)}}{B_2^{(\phi)}} = 
-0.086.
\ee
After expanding (\ref{m4m51}) in 
$\epsilon_1 M$, which is possible as long as $m_5^2$ is not too close 
to $m_2^2$ (so that $M$ is of order one), we find
\be
m_4^2=m_2^2 \, \left[1 + \epsilon_1 \,(\mu^2 -1)\, M(m_5^2) + 
O((\epsilon_1 M)^2) \right ].
\ee
This equation depends only weakly on $m_5^2$. For $m_5^2 > m_3^2$, it 
requires $m_4^2$ to lie very close to the asymptotic value $m_4^2 = 
1.114\, m_2^2$ (where $m_5^2 \rightarrow \infty$) and thus very close 
to the $\phi$ pole and quite far from the next known resonance, the 
$\omega(1420)$. For $m_5^2$ close to the $\phi$, on the other hand, 
the singularity in (\ref{m4m51}) becomes dominant and all values of 
$m_4^2$ can be found as solutions for $m_5^2$ values $ 1.072 \, m_2^2$. 

In other words, one of the two largest pole masses remains always 
close to the $\phi$ mass, whereas the other one is practically 
unconstrained. Qualitatively the same conclusion can be drawn from eq. 
(\ref{m4m52}), which is even more restrictive since $|\epsilon_2| 
< |\epsilon_1|$ and since $\mu$ appears in higher powers. 

Inspection of the solutions for the couplings
\be
B_1^{(v)} &=& - B_1^{(2)} \, \prod_{i =1,3,4,5 \neq v} 
\frac{m_2^2-m_i^2}{m_v^2-m_i^2} , \label{coupl1bis} \\
B_2^{(v)} &=& - B_2^{(2)} \, \frac{m_2^2}{m_v^2}\, 
\prod_{i =1,3,4,5 \neq v} \frac{m_2^2-m_i^2}{m_v^2-m_i^2} ,  
\qquad v \in \{3,4,5\}, \label{coupl2bis} 
\ee
which follow from eqs. (\ref{coupl1}) and (\ref{coupl2}) after the 
additional constraints are implemented, clarifies the significance of 
this result. The above expressions
show that the coupling of the pole close to the $\phi$  
approximately equals the $\phi$ coupling with opposite 
sign, while the remaining two couplings stay small. The large 
contribution of the $\phi$ at asymptotic momenta can therefore
only be canceled (as required by the superconvergence relations)
by a very closely neighboring pole. 

It is worth emphasizing that this result rests on only two qualitative  
features of the spectral density, namely the dominance of the $\phi$ 
coupling over the $\omega$ coupling and the asymptotic decay of the form 
factors. Taking also $m_3$ as a free parameter, in particular, would not 
change the above conclusions, except that in this case one out of ${\it 
three}$ pole masses (i.e. $m_3, m_4$ or $m_5$) would have to stay close 
to $m_\phi$, again with an approximately equal coupling of opposite sign. 

With the help of the solutions (\ref{coupl1bis}) and (\ref{coupl2bis}), 
we can now eliminate the couplings $B_i^{(3)} - B_i^{(5)}$ from the form 
factors, and after rationalizing the results we obtain 
\be
F_1^s(q^2) = q^2 \, \sum_{v=1}^5 B_1^{(v)} m_v^{8} 
\prod_{i=1}^5 \frac{1 }{m^2_i - q^2} = q^2 B_1^{(2)} \frac{ 
\prod_{v \ne 2}^5 (m_v^2-m_2^2)}{ \prod_{v=1}^5 (m_v^2-q^2)} , 
\ee
where $m_4^2$ and $m_5^2$ are related by eq. (\ref{m4m51}), and 
\be
F_2^s(q^2) = \sum_{i=1}^5 \, B_2^{(i)}  \prod_{j=1}^5 
\frac{m^2_j  }{m^2_j - q^2} = B_2^{(2)} m^2_2 \frac{ \prod_{v 
\ne 2}^5 (m_v^2-m_2^2)}{ \prod_{v=1}^5 (m_v^2-q^2)},
\ee
where $m_4^2$ and $m_5^2$ are related by eq. (\ref{m4m52}). 

In order to fix the values of $m_4$ and $m_5$ we now have to search for 
simultaneous solutions of Eqs. (\ref{m4m51}) and (\ref{m4m52}). 
Unfortunately, it is straightforward to see that only trivial solutions 
exist: $m_4^2= m_1^2, m_5^2= m_2^2$ and $m_4^2= m_2^2, 
m_5^2= m_1^2$. (In this case the fourth and fifth pole just cancel the 
$\omega$ and $\phi$ poles and $B_i^{(3)}=0$, i.e. both form factors 
vanish identically.) Thus the 5-pole ansatz with intrinsic asymptotics 
is overconstrained, even if the value of $m_3$ is taken as a free 
parameter. 

Nontrivial approximate solutions of this ansatz do exist, however. 
To see this, note that for 
$\epsilon_2 = \epsilon_1 \mu^2$ the two equations (\ref{m4m51}) and 
(\ref{m4m52}) would coincide and therefore not fix $m_4$ and $m_5$ 
individually. The actual parameter values are not too far from this 
situation ($\epsilon_2 - \epsilon_1 \mu^2 = -0.0458$) and thus 
approximate solutions are possible. The one with the smallest error 
has $m_5^2$ close to the $\phi$ mass and a large value of $m_4^2$. 

Alternatively, one could allow $m_5$ to take slightly different values 
in both form factors. This is not ruled out a priori since the fifth 
pole summarizes the strength of higher lying resonance and continuum 
contributions with potentially  different individual Dirac and Pauli 
couplings. However, none of these approximate solutions resolves  
the problem that neighboring strength in the spectral functions would 
have to balance the $\phi$ pole. This probably makes them unphysical 
since the $\phi$ meson is the only established $0^- 1^{- -}$ resonance 
with a mass in the vicinity of $1 \,{\rm GeV}$ and since there is no 
indication for comparable continuum strength of opposite sign in its 
neighborhood. 

We have also considered a modified 5-pole ansatz with the highest two 
superconvergence relations (which are most sensitive to the higher 
lying spectral strength and thus the least reliable\footnote{Some 
dispersion theoretical studies \cite{que74} have found  that SCRs 
involving higher powers of the masses can overburden the pole 
approximation.}) relaxed, i.e. with $\sum B_1^{(v)} m_v^6 \neq 
0$ and $\sum B_2^{(v)} m_v^8 \neq 0$. This leaves the two masses $m_4$ 
and $m_5$ undetermined and leads to an asymptotic decay intermediate 
between that of the intrinsic and extrinsic contributions. 

The modified 5-pole ansatz leads to $r_2^2$ and $\mu_s$ values in 
the ranges 
\be 
-0.14 \ge \mu_s \ge -0.26, \quad 0.11 \, {\rm fm}^2 \le r_s^2 \le 0.22 
\, {\rm fm}^2 \quad  0.10 \, {\rm fm}^2 \le (r_s^2)_{Sachs} \le 0.20 
\, {\rm fm}^2,  \label{5pbounds}
\ee
for $m_4 \le 1.9 \, {\rm GeV}, m_5 \le 2.2 \,{\rm GeV}$ to 
$m_{4,5} \rightarrow \infty$. Two further poles in the region around 
$2 \,{\rm GeV}$ thus reduce the 3-pole results by up to a factor of two. 
The modified 5-pole form factors with $m_4 =1.9 \, {\rm GeV}, m_5= 2.2 
\,{\rm GeV}$ are well fitted by
\be
F_1^s(q^2) &=& \frac16 \frac{r_s^2 q^2}{( 1-\frac{q^2}{M_1^2} )^4} ,  
\label{dip16p} \\
F_2^s(q^2) &=& \frac{\mu_s}{( 1-\frac{q^2}{M_s^2} )^4},
\label{dip26p}
\ee
with $M_1 = 1.61 {\rm GeV},  M_2 = 1.54 {\rm GeV}$. As expected (and
confirmed by inspection of the higher-pole couplings) all poles except 
the first contribute significantly in this case, and the weighted 
average of their positions corresponds to the values of the mass 
parameters $M_1$ and $M_2$ above.

\begin{table} 
\caption{Fit parameters from Refs.~\protect\cite{hoe76} and 
~\protect\cite{mer95} with $F_1^{I=0}(q^2) = \sum_{v} A_1^{(v)} 
m_v^2 /(m_v^2-q^2)$, $ F_2^{I=0}(q^2) = \sum_{v} A_2^{(v)} m_v^2 
/(m_v^2-q^2)$. For the (weakly momentum-dependent, see section 
~\protect\ref{3polup}) couplings of the MMD fits we list the on-shell 
values.}
\begin{tabular}{|ccccc|} 
Fit &  & & & \\ \hline
8.1 & $m_v^2 \,(\GeV^2)$ & 0.61 & 1.04 & 1.96 \\
 & $A_1^{(v)} $ & 1.13  & -0.52 & -0.11 \\
 & $A_2^{(v)} $ & -0.23 & 0.19  & -0.036  \\
\hline
8.2 & $m_v^2 \, (\GeV^2)$ & 0.61  & 1.04 & 3.24 \\
 & $ A_1^{(v)} $ & 1.17  & -0.62 & -0.040 \\
 & $ A_2^{(v)} $ & -0.18 & 0.13 & -0.0062 \\
\hline
7.1 & $m_v^2\, (\GeV^2)$ & 0.61 & 1.04 & 2.79 \\
 & $ A_1^{(v)} $ & 1.12  & -0.53 & -0.086 \\
 & $ A_2^{(v)} $ & -0.26 & 0.24 & -0.029 \\
\hline
MMD & $m_v^2 \, (\GeV^2)$ & 0.611 & 1.039 & 2.560 \\
 & $ A_1^{(v)} $ & 1.223  & -0.711 & -0.0149 \\
 & $ A_2^{(v)} $ & -0.200 & 0.156  & -0.0159 \\
\end{tabular}
\label{tab-1}
\end{table}

\begin{table}
\caption{Mass and coupling parameters in the 3-pole strange form factors 
$F_1^{(s)} (q^2) = q^2 \sum_{v} B_1^{(v)} / (m_v^2-q^2)$, $ F_2^{(s)} 
(q^2) = \sum_{v} B_2^{(v)} m_v^2 / (m_v^2-q^2)$, as obtained from 
H{\"o}hler et al. \protect\cite{hoe76} (Fits 8.1, 8.2, 7.1) and from 
Mergell, Meissner and Drechsel (MMD) \protect\cite{mer95}.}
\begin{tabular}{|ccccc|}
fit number &  & & & \\
\hline
8.1 & $m_v^2 \, (\GeV^2)$ & 0.61 & 1.04 & 1.96 \\
 & $B_1^{(v)} $ & -0.24  & 1.62 & -1.38 \\
 & $B_2^{(v)} $ & 0.048 & -0.60  & -0.30  \\
\hline
8.2 & $m_v^2 \, (\GeV^2)$ & 0.61  & 1.04 & 3.24 \\
 & $ B_1^{(v)} $ & -0.24  & 1.92 & -1.67 \\
 & $ B_2^{(v)} $ & 0.038 & -0.39 & 0.12 \\
\hline
7.1 & $m_v^2 \, (\GeV^2)$ & 0.61 & 1.04 & 2.79 \\
 & $ B_1^{(v)} $ & -0.23  & 1.65 & -1.41 \\
 & $ B_2^{(v)} $ & 0.055 & -0.75 & 0.27 \\
\hline
MMD & $m_v^2 \, (\GeV^2)$ & 0.611 & 1.039 & 2.560 \\
 & $ B_1^{(v)} $ & -0.256  & 2.214 & -1.958 \\
 & $ B_2^{(v)} $ & 0.0420 & -0.486 & 0.187 \\
\end{tabular}
\label{tab-2}
\end{table}

\begin{table}
\caption{Theoretical estimates of the strangeness radius and magnetic 
moment}
\begin{tabular}{|c|c|c|c|c|}
 Source & $\mu_s \,\, (\mu_N) $ & $r^2_s \,\, (10^{-2} {\rm \fm}^2)$ & 
$(r^2_s)_{Sachs} \,\, (10^{-2} {\rm \fm}^2)$ & Ref. \\ 
\hline 
3 poles  &-- 0.31 $\pm 0.09$ & 16 & 14 $\pm$ 0.07 & \cite{jaf89}\\ 
3 poles, updt.  & -- 0.26  & 22 & 20 & this work \\ 
4 poles, extr.  & -- (0.18 -- 0.26)  & 15 -- 22 & 14 -- 20 & this work \\ 
6 poles, intr.  & -- (0.11 -- 0.26)  & 8.9 -- 22 & 8.1 -- 20 & this work \\ 
Kaon-loops \& SU(3) & -- (0.31 -- 0.40)  & -- (0.67 -- 0.59) & 
-- (2.71 -- 3.23) & \cite{mus93} \\  
Kaon-loops & -- (0.24 -- 0.32)  & -- (0.67 -- 0.69) & -- (2.23 -- 2.76) & 
\cite{for94,nos}\\  
Gen. VMD & -- (0.24 -- 0.32)  & -- (2.43 -- 2.45) & -- (3.99 -- 4.51) & 
\cite{for94,nos} \\ 
CQ-NJL &  &  & 1.69 & \cite{for94} \\
SU(3) Skyrme & -- 0.13 & -- 10 & -- 11 & \cite{par91} \\
Bos. NJL I & -- (0.05 -- 0.25) & -- (10 -- 20) & -- (10 - 22) & 
\cite{wei92} \\
Bos. NJL II & -- 0.45 & -- 17 & -- 20 & \cite{kim95} \\
\end{tabular}
\label{tab-3}
\end{table}

\newpage

\begin{figure}
{\footnotesize FIG. 1a. The strange Dirac form factor from the 3-pole 
ansatz with parameters (cf. Table II) based on Fit 8.1 (dashed), Fit 7.1 
(dot-dashed), and Fit 8.2 (dotted) of H{\"o}hler et al. 
~\protect\cite{hoe76}, and on the fit of MMD ~\protect\cite{mer95} 
(full line).}
\label{fig-1}
\end{figure}
\begin{figure}
{\footnotesize FIG 1b. The strange Pauli form factor from the 3-pole 
ansatz with lines defined as in Fig. 1.}
\label{fig-2}
\end{figure}
\begin{figure}
{\footnotesize FIG 2a. The strange Dirac form factor from the 3-pole 
(full line) ansatz, the 4-pole ansatz with extrinsic asymptotics (dotted 
line) and the 6-pole ansatz with intrinsic asymptotics (dashed line). }
\label{fig-3}
\end{figure}
\begin{figure}
{\footnotesize FIG 2b. The strange Pauli form factor from the 3-pole, 
4-pole and 6-pole ans{\"a}tze in the same notation as in FIG 2a.}
\label{fig-4}
\end{figure}


\begin{references}

\bibitem{der75} A. DeRujula, H. Georgi and S.L. Glashow, Phys. 
Rev. D 12, 147 (1975); N. Isgur and G. Karl, Phys. Rev. D 18, 
4187 (1978).

\bibitem{bro81} S.J. Brodsky, C. Peterson and N. Sakay, Phys. Rev. 
D23, 2745 (1981).

\bibitem{che71}T.P. Cheng and R. Dashen, Phys. Rev. Lett. 26, 
594 (1971); T.P. Cheng, Phys. Rev. D 13, 2161 (1976).

\bibitem{don86}J.F. Donoghue and C.R. Nappi, Phys.\ 
Lett.\ B168, 105 (1986). 

\bibitem{gas91} J. Gasser, H. Leutwyler and M.E. Sainio, Phys.\ 
Lett.\ B253, 252 (1991).

\bibitem{ahr87} L.A. Ahrens et al., Phys.\ Rev.\ D 35, 785 
(1987).

\bibitem{ash88}J. Ashman et al., Phys.\ Lett.\  B206, 364 (1988); 
Nucl. Phys. B328, 1 (1989). 

\bibitem{deut} P. Amaundruz et al., Phys. Rev. Lett. 66, 2712 
(1991); CERN-PPE/93-117; B. Adeva et al., Phys. Lett. B 302, 
533 (1993); Phys. Lett. B 320, 400 (1994).

\bibitem{iof91} see, for example, B.L. Ioffe and M. Karliner, 
Phys. Lett. B 247, 387 (1990).

\bibitem{gar93} G.T. Garvey, W.C. Louis and D.H. White, 
Phys. Rev. C 48, 761 (1993).

\bibitem{mck89} R.D. McKeown, Phys.\ Lett.\ B219, 140 (1989).

\bibitem{bec89} D.H. Beck,  Phys. Rev. D 39, 3248 (1989).

\bibitem{mus94} M.J. Musolf, T.W. Donnelly, J. Dubach, 
S.J. Pollock, S. Kowalski and E.J. Beise, Phys. Rep. 239, 1 
(1994).

\bibitem{ce17}CEBAF proposal \#PR-91-017, D.H. Beck, spokesperson.

\bibitem{ce4}CEBAF proposal \#PR-91-004, E.J. Beise, spokesperson.

\bibitem{ce10}CEBAF proposal \#PR-91-010, J.M. Finn and P.A. 
Souder, spokespeople.

\bibitem{hei93} MAMI proposal A4/1-93, D. von Harrach, spokesperson.

\bibitem{mit}MIT/Bates proposal \# 89-06, R. McKeon and D.H.
Beck, contact people.

\bibitem{par91} N.W. Park, J. Schechter and H. Weigel, Phys.\ Rev.\ 
D 43, 869 (1991); N.W. Park and H. Weigel, Nucl.\ Phys.\ A541, 
453 (1992).

\bibitem{wei92} H. Weigel, R. Alkofer and H. Reinhardt, Nucl. Phys. 
B387, 638 (1992).

\bibitem{mus93} M.J. Musolf and M. Burkardt, Z. Phys. C61, 433 
(1994).

\bibitem{nos} T.D. Cohen, H. Forkel and M. Nielsen, Phys. Lett. 
B316, 1 (1993).

\bibitem{hon93} S. Hong and B. Park, Nucl. Phys. A561, 525 (1993).

\bibitem{for94} H. Forkel, M. Nielsen, X. Jin, and T.D. Cohen, 
Phys. Rev. C 50, 3108 (1994).

\bibitem{koe94} W. Koepf and E.M. Henley, Phys.\ Rev.\ 
C 49, 2219 (1994).

\bibitem{kim95} H.-C. Kim, T. Watabe and K. Goeke, preprint 
RUB-TPII-11/95.
\bibitem{keh94} S.J. Dong and K.F. Liu, Phys. Lett. B328, 
130 (1994).

\bibitem{jaf89} R.L. Jaffe, Phys.\ Lett.\ B229, 275 (1989).

\bibitem{for95} H. Forkel, Plenary talk at the International 
School for Nuclear Physics, Erice, 17th Course:``Quarks in 
Hadrons and Nuclei'', Sept. 19-27, 1995, Prog. Part. Nucl. Phys. 
{\bf 36}, 229 (1996). 

\bibitem{hoe76} G. H{\"o}hler et al., Nucl.\ Phys.\ B114, 505 
(1976).

\bibitem{mer95} P. Mergell, U.-G. Meissner and D. Drechsel,
Mainz and Bonn University preprint TK 9515, MKPH-T-95-07,
hep-ph/9506375

\bibitem{hoe93} G. H{\"o}hler, $\pi-N$ Newsletter 9, 108 (1993).

\bibitem{jain}P. Jain et al., Phys.\ Rev.\ D 37, 3252 (1988). 

\bibitem{bro75} S.J. Brodsky and G.R. Farrar, Phys. Rev. D11, 1309
(1975).

\bibitem{lep80} S.J. Brodsky and G.P. Lepage, Phys. Rev. D22, 2157
(1980).

\bibitem{sab75} S. Ciulli, C. Pomponiu and I. Sabba Stefanescu, 
Phys. Rep. 17, 133 (1975); I. Sabba Stefanescu, J. Math. Phys. 
21, 175 (1980).

\bibitem{gen76} H. Genz and G. H{\"o}hler, Phys. Lett. B61, 
389 (1976).

\bibitem{gar92} M.F. Gari and W. Kr"umpelmann, Phys. Lett. B274, 
159 (1992).

\bibitem{gal71} S. Galster et al., Nucl. Phys. B32, 221 (1971).

\bibitem{for96} H. Forkel, in preparation 

\bibitem{dub92} S. Dubnicka, A.Z. Dubnickova and P. Strizenec,
Dubna preprint E2-92-520 (1992); S.I. Bilenkaya et al., Nouvo
Cim. A 105 (1992) 1421.

\bibitem{for952} H. Forkel, M.J. Musolf and M. Nielsen, in 
preparation

\bibitem{que74} N.M. Queen and G. Violini, Dispersion Theory in 
High-Energy Physics, John Wiley, New York, 1974.


\end{references}
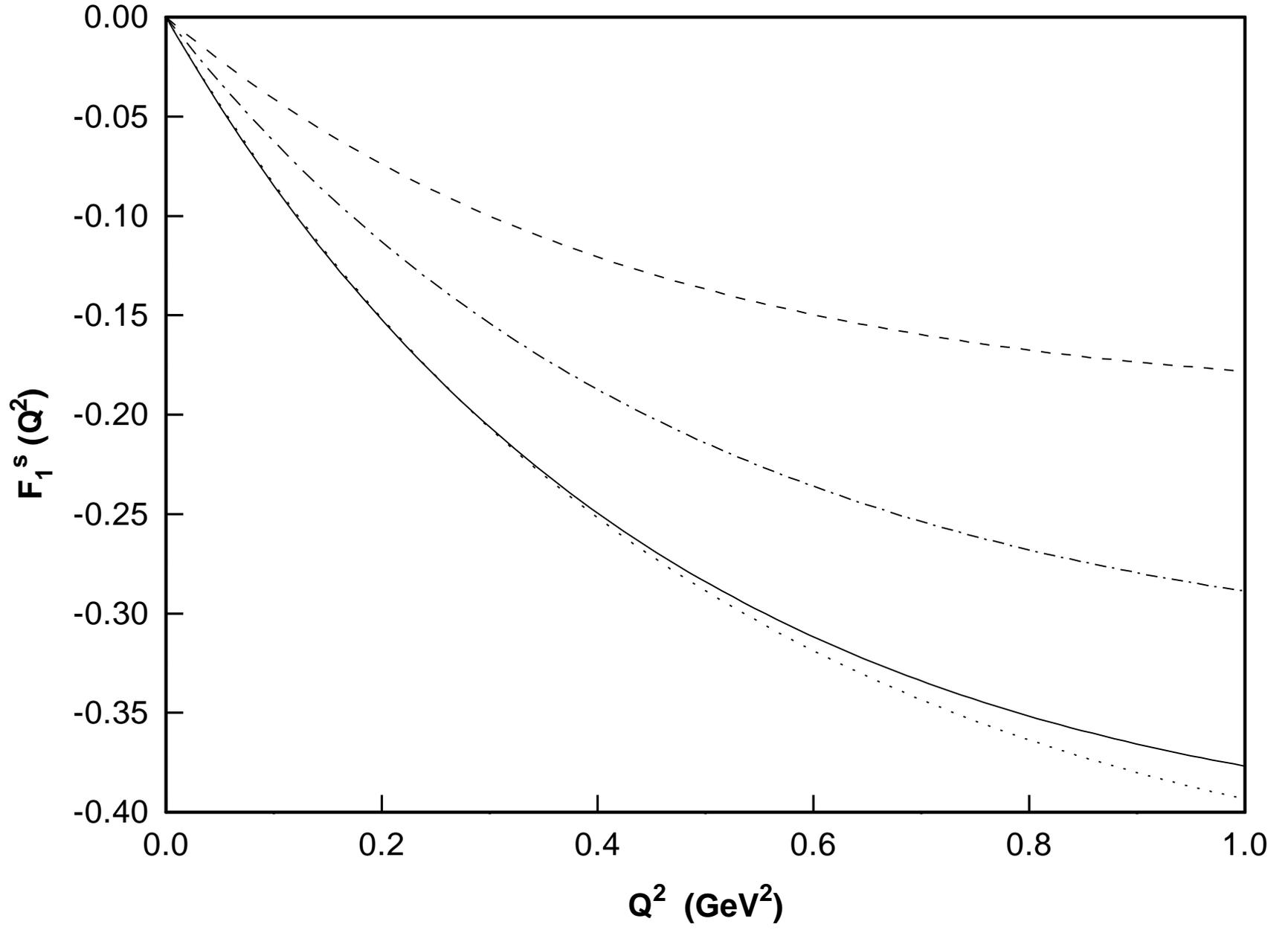
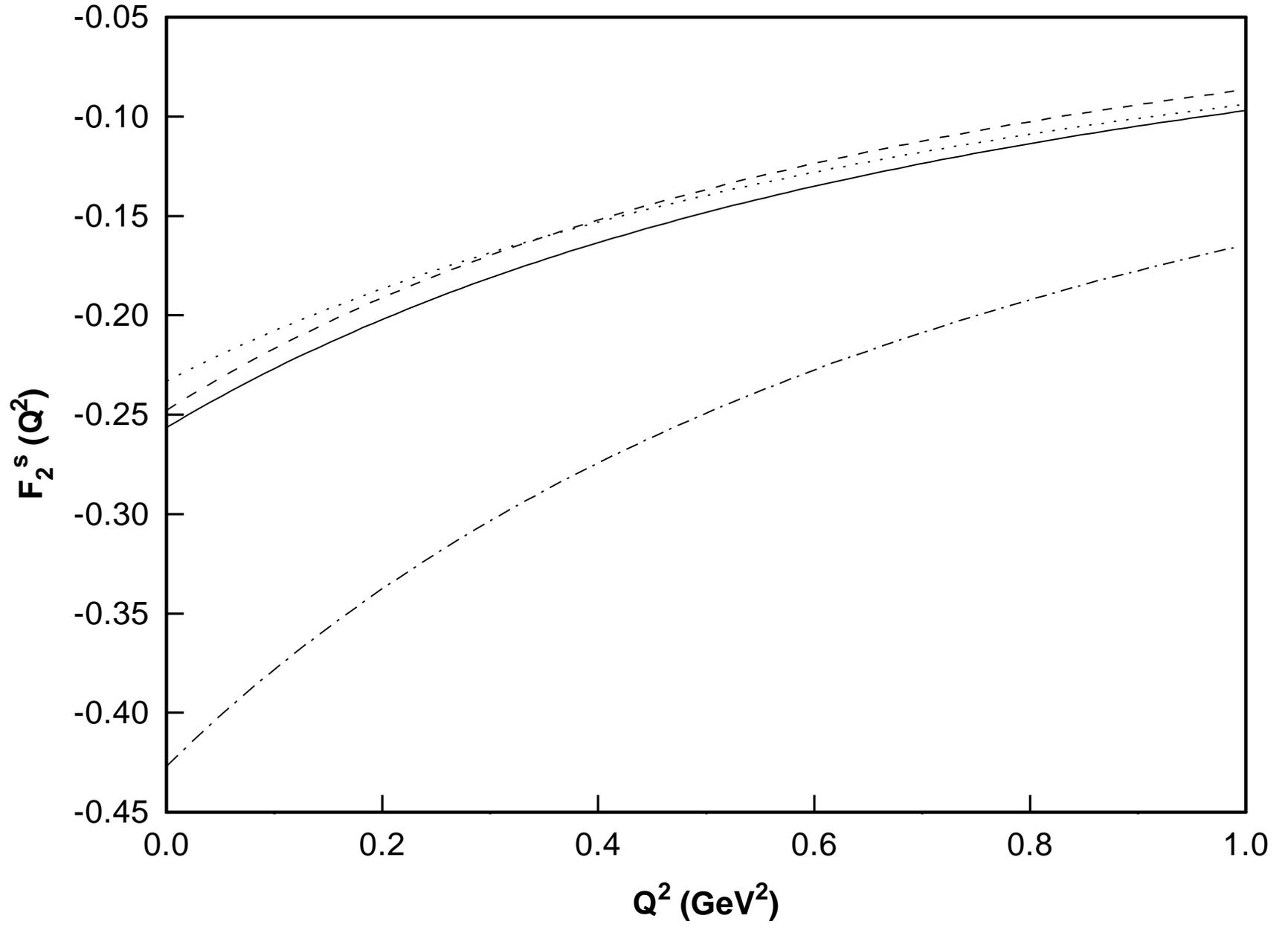
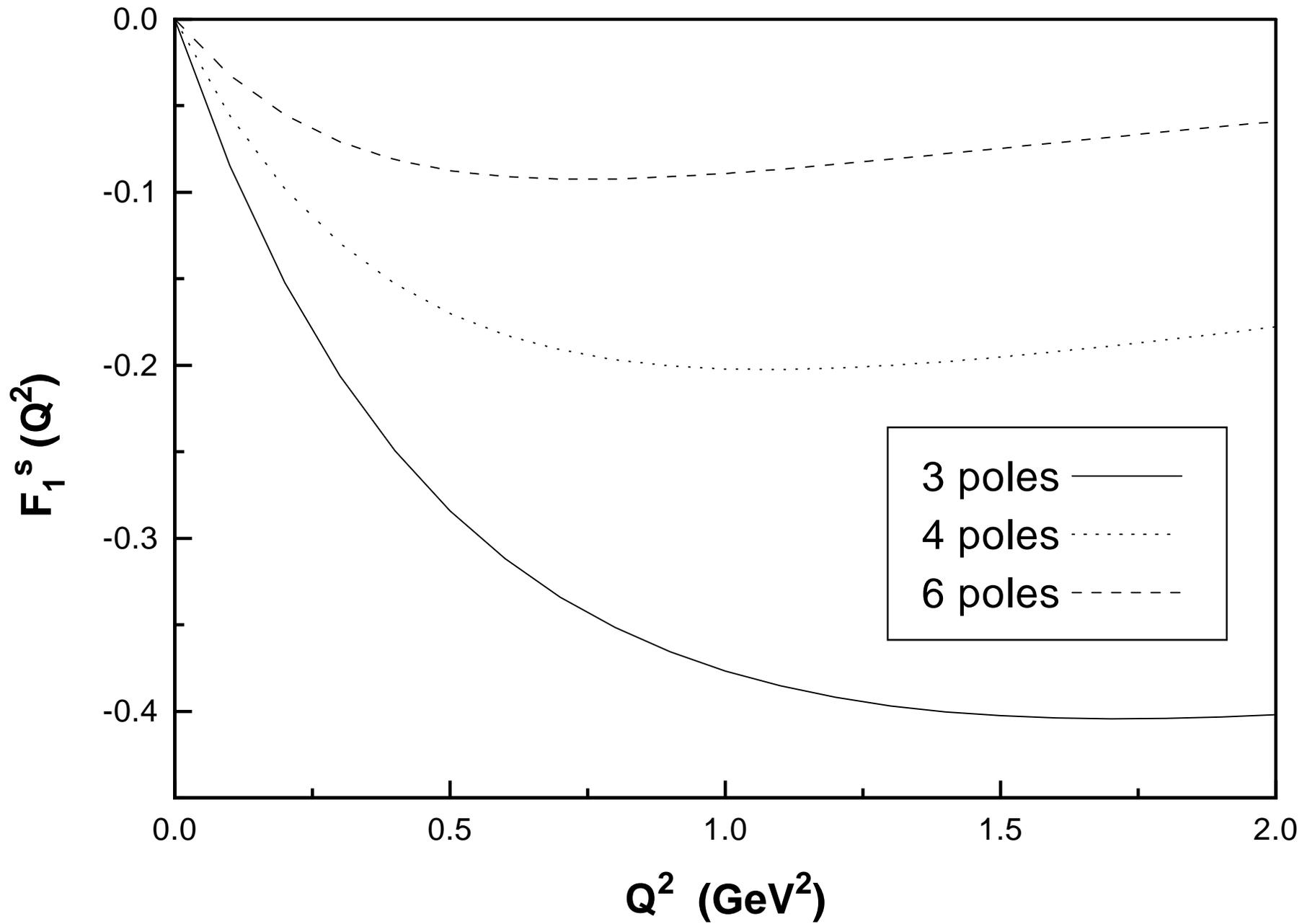
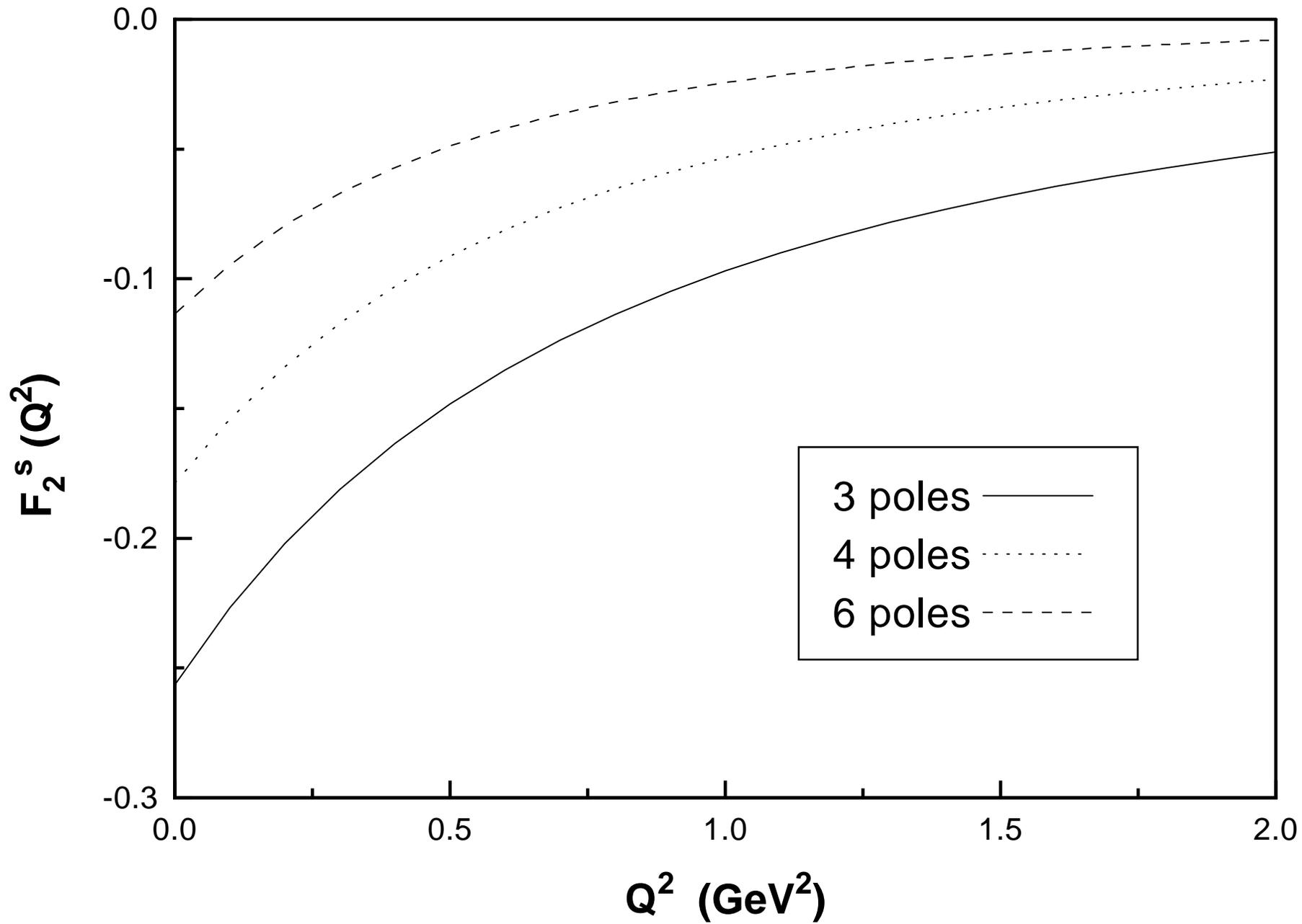
\end{document}